\documentclass[twocolumn]{aastex63}

\newcommand{\kms}{$\mathrm{km\,s^{-1}}$}
\newcommand{\feh}{$\rm [Fe/H]$}
\newcommand{\Msun}{$\rm M_{\odot}$}
\newcommand{\Lsun}{$\rm L_{\odot}$}
\newcommand{\MLsun}{$\rm M_{\odot}/L_{\odot}$}

\usepackage{enumerate}

\def\tablefoot#1{\par\vspace*{2ex}%
 \parbox{\hsize}{\leftskip0pt\rightskip0pt
 {\noindent\small\textbf{Notes.}~#1\par}}}
\def\tablefootmark#1{$^{#1}$\,\ignorespaces}
\def\tablefoottext#1#2{$^{(#1)}$~#2}

\received{\today} %{June 1, 2019}
\revised{\today} %{January 10, 2019}
\accepted{\today}
\submitjournal{ApJ}

\shorttitle{M\,54 population-dynamical model}
\shortauthors{Kacharov et al.}
\begin{document}

\title{A Deep View into the Nucleus of the Sagittarius Dwarf Spheroidal Galaxy with MUSE. \\
	III. Discrete multi-component population-dynamical models based on the Jeans equations}

\correspondingauthor{Nikolay Kacharov}
\email{kacharov@aip.de}

\author{Nikolay Kacharov}
%\affiliation{Max Planck Institute for Astronomy, K\"onigstuhl 17, 69117 Heidelberg, Germany}
\affiliation{Leibniz Institute for Astrophysics, An der Sternwarte 16, 14482 Potsdam, Germany}

\author{Mayte Alfaro-Cuello}
%\affiliation{Max Planck Institute for Astronomy, K\"onigstuhl 17, 69117 Heidelberg, Germany}
\affiliation{Space Telescope Science Institute, 3700 San Martin Drive, Baltimore, MD 21218, USA}

\author{Nadine Neumayer}
\affiliation{Max Planck Institute for Astronomy, K\"onigstuhl 17, 69117 Heidelberg, Germany}

\author[0000-0002-4034-0080]{Nora L\"utzgendorf}
\affiliation{European Space Agency (ESA), ESA Office, Space Telescope Science Institute, 3700 San Martin Drive, Baltimore MD 21218, USA}

\author[0000-0002-1343-134X]{Laura L. Watkins}
\affiliation{AURA for the European Space Agency, ESA Office, Space Telescope Science Institute, 3700 San Martin Drive, Baltimore MD 21218, USA}

\author{Alessandra Mastrobuono-Battisti}
\affiliation{GEPI, Observatoire de Paris, PSL Research University, CNRS, Place Jules Janssen, 92190 Meudon, France}

\author{Sebastian Kamann}
\affiliation{Astrophysics Research Institute, Liverpool John Moores University, 146 Brownlow Hill, Liverpool L3 5RF, UK}

\author{Glenn van de Ven}
\affiliation{Department of Astrophysics, University of Vienna, T\"urkenschanzstrasse 17, 1180 Wien, Austria}

\author{Anil C. Seth}
\affiliation{Department of Physics and Astronomy, University of Utah, Salt Lake City, UT 84112, USA}

\author{Karina T. Voggel}
\affiliation{Universit\'e de Strasbourg, CNRS, Observatoire astronomique de Strasbourg, UMR 7550, F-67000 Strasbourg, France}

\author{Iskren Y. Georgiev}
\affiliation{Max Planck Institute for Astronomy, K\"onigstuhl 17, 69117 Heidelberg, Germany}

\author{Ryan Leaman}
\affiliation{Max Planck Institute for Astronomy, K\"onigstuhl 17, 69117 Heidelberg, Germany}
\affiliation{Department of Astrophysics, University of Vienna, T\"urkenschanzstrasse 17, 1180 Wien, Austria}

\author{Paolo Bianchini}
\affiliation{Universit\'e de Strasbourg, CNRS, Observatoire astronomique de Strasbourg, UMR 7550, F-67000 Strasbourg, France}

\author[0000-0002-5666-7782]{Torsten B\"oker}
\affiliation{European Space Agency (ESA), ESA Office, Space Telescope Science Institute, 3700 San Martin Drive, Baltimore MD 21218, USA}

\author{Steffen Mieske}
\affiliation{European Southern Observatory, 3107 Alonso de C\'ordova, Vitacura, Santiago, Chile}



\begin{abstract}

We present comprehensive multi-component dynamical models of M\,54 (NGC\,6715), the nuclear star cluster of the Sagittarius dwarf galaxy (Sgr), which is undergoing a tidal disruption in the Milky Way halo.
Previous papers in the series used a large MUSE mosaic data set to identify multiple stellar populations in the system and study their kinematic differences.
Here we use Jeans-based dynamical models that fit the population properties (mean age and metallicity), spatial distributions, and kinematics simultaneously.
They provide a solid physical explanation to our previous findings.
The population-dynamical models deliver a comprehensive view of the whole system, and allow us to disentangle the different stellar populations.
We explore their dynamical interplay and confirm our previous findings about the build-up of Sgr's nuclear cluster via contributions from globular cluster stars, Sgr inner field stars, and in-situ star formation.
We explore various parameterisations of the gravitational potential and show the importance of a radially varying mass-to-light ratio for the proper treatment of the mass profile.
We find a total dynamical mass within M\,54's tidal radius ($\sim75$\,pc) of $1.60\pm0.07\times10^6$\,\Msun in excellent agreement with $N$-body simulations.
The metal-poor globular cluster stars contribute about $65\%$ of the total mass or $1.04\pm0.05\times10^6$\,\Msun.
The metal-rich stars can be further divided into young and intermediate age populations that contribute $0.32\pm0.02\times10^6$\,\Msun ($20\%$) and $0.24\pm0.02\times10^6$\,\Msun ($15\%$), respectively.
Our population-dynamical models successfully distinguish the different stellar populations in Sgr's nucleus because of their different spatial distributions, ages, metallicities, and kinematic features.

\end{abstract}



\section{Introduction} \label{sec:intro}

Nuclear star clusters (NSC) are common among low- and intermediate-mass galaxies of all morphological types, including dwarf galaxies with stellar masses lower than $10^9$\,\Msun~\citep[see recent review by][]{neumayer+2020}.
Their extreme stellar densities \citep[$\rm >10^6\,M_{\odot}/pc^3$][]{hopkins+2010}, possible coincidence with supermassive black holes \citep{filippenko+ho2003,seth+2008,graham+spitler2009,neumayer+walcher2012,nguyen+2019}, and extended star formation histories with coexistence of multiple stellar populations \citep{walcher+2006,rossa+2006,norris+2015,kacharov+2018,fahrion+2021,hannah+2021} make such objects very interesting from a population-dynamical point of view.

The Sagittarius (Sgr) dwarf spheroidal galaxy (dSph), discovered by \citet{ibata+1994} is the closest known example of a nucleated dwarf galaxy \citep{bellazzini+2008}.
It is currently being tidally stripped by the Milky Way, but its nucleus - the complex, massive, and dense star cluster M\,54 is still largely intact \citep{bassino+1994, bekki+2003, pfeffer+baumgardt2013, pfeffer+2021}.

This work continues the series of papers on the Sgr dSph NSC - M\,54, based on VLT MUSE observations by \citet{paper1} and \citet{paper2}, hereafter Paper~I \& Paper~II.
In Paper~I we introduced our observational sample and classified M\,54's stars in three different stellar populations, based on measured metallicities and ages.
We named these populations old metal-poor (OMP), intermediate-age metal-rich (IMR), and young metal-rich (YMR).
We found out that the YMR population is the most centrally concentrated and the IMR is the most spatially extended.
In Paper~II we explored their kinematic properties and showed that they also differ considerably.

Sgr's nuclear formation picture that emerged from these studies is in agreement with the most widely accepted scenarios for NSC formation, namely that both proposed mechanisms - {\em in situ} star formation from accreted gas \citep{mihos+hernquist1994,milosavljevic2004,schinnerer+2008,bekki2015} and mergers of globular clusters \citep{tremaine+1975,oh+lin2000,lotz+2001,capuzzo-dolcetta+miocchi2008a,capuzzo-dolcetta+miocchi2008b,antonini+2012,gnedin+2014} operate simultaneously \citep{neumayer+2011,denBrok+2014,antonini+2015,cole+2017,fahrion+2022}.

In the third instalment of the paper series we focus on constraining the over-all shape of the gravitational potential and explore the interplay between the various distinct M\,54 stellar populations in the context of NSC formation theories.

The OMP population old age and kinematic properties are typical for globular clusters.
They show negligible rotation and a radially decreasing velocity dispersion profile.
Its spatial distribution can be well described with a King profile \citep{king1962, trager+1995, mclaughlin+vandermarel2005, monaco+2005}.
Its significant metallicity spread \citep[$0.24$\,dex, Paper~I; $0.19$\,dex][]{carretta+2010}, compared to the typical globular cluster metallicity spread of $<0.1$\,dex \citep{carretta+2009}, suggests that the OMP stars might be the result of multiple merged globular clusters that have sunk to Sgr's centre via dynamical friction.
In addition, stars from the metal-poor end of the Sgr field distribution likely also co-exist in the nuclear cluster, inflating M\,54's OMP metallicity spread.

The YMR stars, on the other hand, exhibit a significantly higher rate of systemic rotation of $\sim5$\,\kms and have a lower velocity dispersion than the OMP stars.
This is expected if they formed {\em in situ} within a high angular momentum disk structure from dynamically cold gas that was accreted in Sgr's central region.
Their mean age of $\sim2$\,Gyr coincides with Sgr's first peri-galactic passage, which could have triggered the nuclear star formation episode \citep{tepper+bland2018, diCintio+2021}.

The IMR population has a significant age and metallicity spread and its star formation history (SFH) and the recovered age-metallicity relation (Paper~I) follows that of Sgr's field population closely \citep{layden+sarajedini2000}.
This population is also characterised by a radially flatter velocity dispersion profile with higher dispersion in the outer nuclear region, compared to the OMP stars.
It is much more extended than the YMR and OMP populations, however it still forms a density cusp, making it distinct from the field stars spatial distribution \citep{monaco+2005}.

Previous dynamical studies of M\,54 have looked into the possibility of detecting an intermediate mass black hole in the centre of the system \citep{ibata+2009, wrobel+2011}, but have not reached conclusive results.
Although our MUSE data is suitable to explore this question with improved sensitivity, we leave it for a future publication.
In addition, as an integral part of the dark-matter dominated dSph galaxy, Sgr's nucleus is also likely embedded in a dark matter halo, which contribution to the gravitational potential starts to become important in the outer regions \citep[$>30$\,pc, e.g.][]{carlberg+2022}.

In this work we base our dynamical models on the \citet{jeans1915} hydrodynamical equations of stellar motions, which are often used today due to their simplicity and computational efficiency \citep{cappellari2008}.
In the past, direct applications of this approach often required binning the data to infer the higher velocity moments, which significantly limited our ability to model complex stellar systems.
\citet{watkins+2013} presented a method to use the Jeans dynamical models with discrete likelihood functions, which takes full advantage of the kinematic measurements for individual stars with their respective uncertainties and allows for the simultaneous modelling of multiple stellar populations \citep{zhu+2016, kamann+2020}.
Our dynamical models explore different parameterisations of the gravitational potential and are based on two or three distinct stellar populations, separated by metallicity and age.

The article is organised as follows:
Sect. \ref{sec:obs} describes the MUSE dataset and briefly summarises the relevant results of the first two papers in this series;
Sect. \ref{sec:dyn} describes and presents the results of our discrete population-dynamical models;
Sect. \ref{sec:discussion} discusses the results in the context of previous work, and Sect. \ref{sec:summary} summarises our findings.

For consistency, we take the base parameters for M\,54 from the Harris catalogue of globular clusters \citep[2010 edition]{harris1996}.
They are summarised in Table \ref{tab:param}.

\begin{table}
	\centering
	\caption{M\,54 adopted parameters}
	\label{tab:param}
	%\resizebox{\columnwidth}{!}{
	\begin{tabular}{lc}
		\hline
	Parameter & Value\tablefootmark{1}  \\
		\hline   
	RA    & $283.76387^{\circ}$  \\
	DEC & $-30.479861^{\circ}$ \\
	$\rm PM_{RA}$ & $-2.683$\,mas/yr \\
	$\rm PM_{DEC}$ & $-1.385$\,mas/yr \\
	Distance & 26.5\,kpc  \\          %Distance & 28.4\,kpc & \citet{siegel+2011} \\
	Core radius ($\rm r_c$) & $5.4\arcsec$  \\
	Half-light radius ($\rm r_h$) & $49.2\arcsec$  \\
	Tidal radius\tablefootmark{2} ($\rm r_t$) & $590\arcsec$  \\
	Abs. V-band mag. & $-9.98$\,mag  \\
	Luminosity ($\rm L_V$) & $0.85\times10^6$\,\Lsun \\
		\hline
	MUSE $\rm r_{max}$ \tablefootmark{3} & $149\arcsec$ \\
	Angular scale & $0.13$\,pc$/$arcsec\\
		\hline
	\end{tabular}
	%}
	\tablefoot{
	\tablefoottext{1}{All values are taken from the Harris catalogue of globular clusters \citep[][ed. 2010]{harris1996}, besides the proper motions, which come from \citet{vasiliev+baumgardt2021}.}
	\tablefoottext{2}{The concept of a tidal radius is not really meaningful in nuclear clusters. In any case, we give our total mass estimates at this radius.}
	\tablefoottext{3}{Diagonal extent of the MUSE mosaic field of view.}
	}
\end{table}

\section{Observations}\label{sec:obs}

In this section, we present the three datasets from the Multi-Unit Spectroscopic Explorer \citep[MUSE][]{bacon+2014} that we use to extract individual stellar spectra and create the M\,54 stellar catalogue.
MUSE is located at UT\,4 (Yepun) of the Very Large Telescope (VLT) at the Paranal Observatory in Chile.

\subsection{MUSE wide field mode}

\begin{figure*}
\centering
\includegraphics[width=500px]{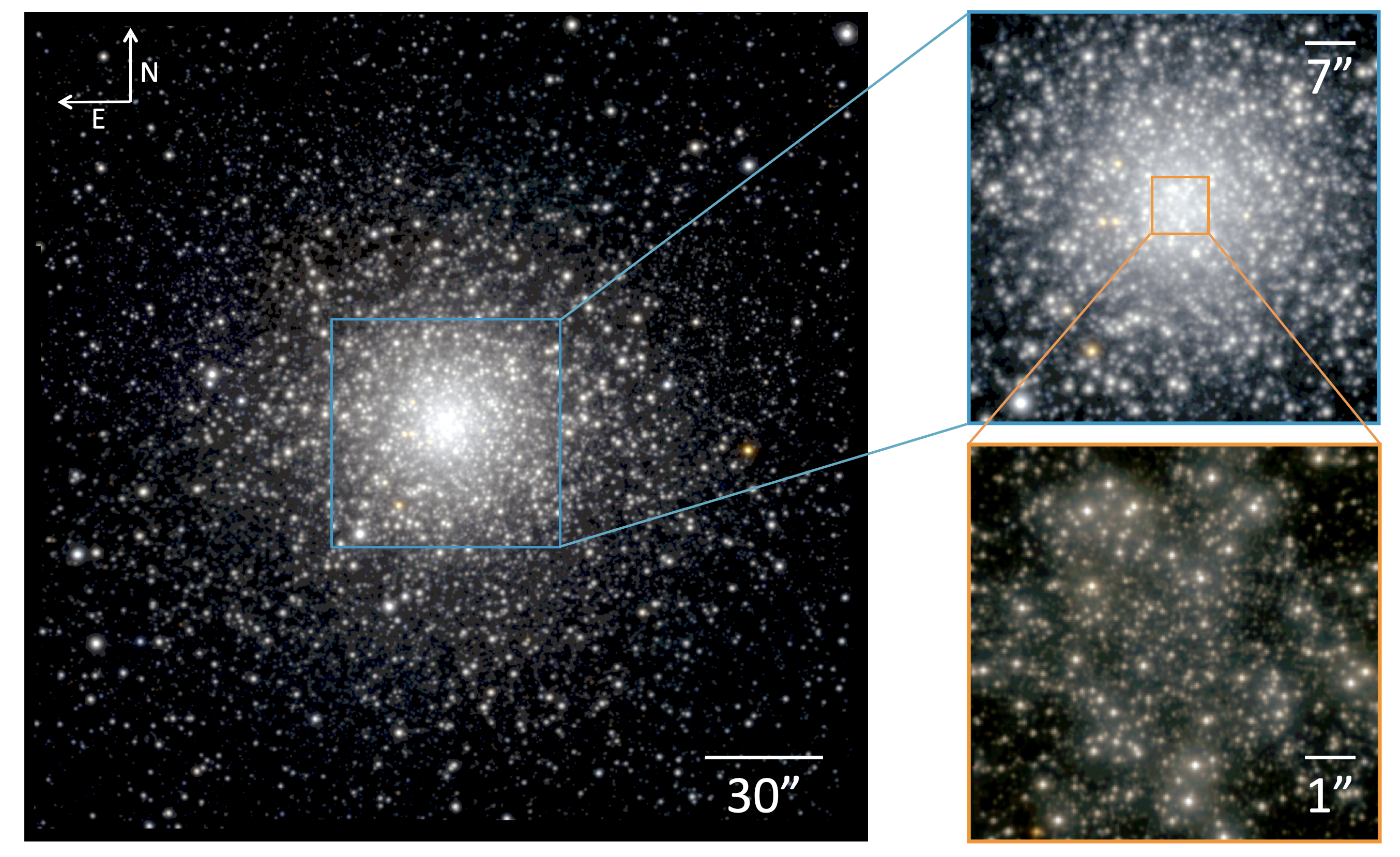}
\caption{Colour images obtained from the MUSE cubes using synthetic {\it i}, {\it r}, and {\it z} filters. Left panel shows the WFM MUSE data composed by a 4$\times$4 mosaic of 16 pointings, covering an area of $\sim$2.5\,$\rm r_h$ (the same as in Paper~I; $\rm r_h=49\arcsec$, \citealt[][2010 edition]{harris1996}).
The blue square shows the position of the single WFM+AO $1'\times1'$ pointing. The WFM+AO colour image is presented in the top right panel. The orange square shows the position of the single NFM pointing, which colour image is presented in the bottom right panel with a corresponding size of $7.5\arcsec\times7.5\arcsec$. North is up and east is to the left.}
\label{fig:muse_mosaic}
\end{figure*}

One part of the dataset used to analyse the dynamical properties of M\,54 consists of our MUSE mosaic observations of the system taken between June 29th and September 19th 2015 during run 095.B-0585(A) (P.I.: L\"utzgendorf).
These observations and the data reduction are described in depth in Paper~I \& Paper~II.
Here we briefly lay out the more important details.
We have a $4\times4$ pointing MUSE wide field mode (WFM) seeing-limited mosaic, centred on M\,54.
The field of view of a single MUSE pointing is $59.9\arcsec\times60.0\arcsec$ with a wavelength coverage in the range of 4800\,-\,9300\,\AA, and a spatial sampling of $0.2\arcsec$\,pix$^{-1}$.
Due to some overlap between the individual pointings, our data covers a square field of view with a side of $210$\arcsec ($27.3$\,pc, assuming a $26.5$\,kpc distance; see Table \ref{tab:param}).
Figure \ref{fig:muse_mosaic} shows a synthetic colour image of the MUSE mosaic, constructed from the MUSE IFU cubes.
According to the catalogue of GCs of \citet[][2010 edition]{harris1996}, M\,54 has a core radius $r_c=5.4\arcsec$ ($0.7$\,pc), a half-light radius $r_h=49\arcsec$ ($6.4$\,pc), and a tidal radius $r_t=592\arcsec$ ($77$\,pc).
Thus, our data covers the system out to $\gtrsim2$ half-light radii.

\subsection{MUSE WFM with Adaptive Optics}

We observed adaptive optics (AO) corrected WFM pointing (WFM+AO), centred on M\,54 on August 8th, 2017 with the MUSE AO facility, which utilises the GALACSI AO module for ground layer atmospheric turbulence correction, as part of the MUSE WFM+AO science verification programme 60.A-9181(A) (PI: Alfaro-Cuello).
The WFM+AO configuration is the same as the seeing limited WFM observations (the same wavelength coverage and spectral and spatial resolution), however due to the sodium lasers used by the AO facility that saturate the detector, there is a gap in the spectra between $5780$ and $6050$\,\AA.
The point spread function (PSF) full width half maximum (FWHM) of the WFM+AO field is $0.6\arcsec$.
We took four science exposures  (750 seconds each), applying $90^{\circ}$ field rotation between them without dithering.
We show the colour image obtained from the WFM+AO data cube using synthetic {\it i}, {\it r}, and {\it z} filters in the top right panel of Figure~\ref{fig:muse_mosaic}.

\subsection{MUSE narrow field mode with AO}

We also include in this analysis a central narrow field mode (NFM+AO) data cube in the very centre of M\,54, taken during the MUSE\,/\,NFM science verification observations on September 8th, 2018 \citep[program ID 60.A-9486(A), P.I. Alfaro-Cuello, first results shown in][]{leibundgut+2019}.
MUSE NFM+AO mode utilises a laser tomographic AO system that provides a $7.42\arcsec \times 7.43\arcsec$ field of view and nearly diffraction limited images.
This single central pointing covers the innermost region of M\,54 corresponding to roughly $1\times1$\,pc at the assumed distance of $26.5$\,kpc. 
The PSF FWHM is better than $0.2\arcsec$.
We present the colour image obtained from the NFM cube using synthetic {\it i}, {\it r}, and {\it z} filters in the bottom right panel of Figure~\ref{fig:muse_mosaic}.
The MUSE NFM mode has the same wavelength coverage as the WFM (4800 to 9300\,\AA), but with a 10 times better spatial sampling of $0.025\arcsec$\,pix$^{-1}$.
Similar to the WFM+AO observations, there is a gap in the spectra between 5780 and 6050\,\AA due to light contamination from the sodium lasers.
We acquired four science exposures (900\,s each) and no rotation nor dithering was applied between exposures.

This latter data set is especially important to resolve individual stellar spectra in the innermost region of the cluster and will be crucial to constrain the central mass profile and for a possible detection of an intermediate mass black hole in the centre of Sgr's NSC, where crowding severely limits the non-AO data (Alfaro-Cuello et al., in prep.).

\subsection{Individual Stellar Spectra Extraction and analysis}

The stellar spectra of resolved stars were extracted with {\sc pampelmuse} \citep{kamann+2013}\footnote{\url{https://pampelmuse.readthedocs.io}}.
This software models the change in the point spread function with wavelength, allowing to de-blend and extract the spectra of sources efficiently even in crowded and dense regions.
This program needs a photometric reference catalogue, for which we use Hubble Space Telescope (HST) photometry from \citet{siegel+2007}.
However, as we pointed out in Paper~II, even with {\sc pampelmuse} we were not able to fully resolve the spectra of the innermost stars, which led to an artificial drop of the velocity dispersion in the central region.
This problem has been significantly alleviated by including the MUSE WFM-AO and NFM observations.

During the stellar spectra extraction with {\sc pampelmuse} we model the PSF using a Moffat profile for the WFM and WFM+AO MUSE datasets.
For the NFM dataset, we use the MAOPPY PSF model profile \citep{fetick+2019} that is better suited better for AO observations and especially the MUSE NFM cubes.

Our final stellar catalogue in the field of view of M\,54 contains $8927$ entries with measured radial velocities and metallicities from spectra with signal-to-noise ratios better than $10$\,px$^{-1}$.
Of these, 3008 sources are extracted from the WFM-AO cube and 627 from the NFM cube.
We analysed all spectra from the three data cubes in a uniform manner with the software tool  {\sc spexxy}\footnote{\url{https://spexxy.readthedocs.io}}\footnote{\url{https://github.com/thusser/spexxy}} \citep{husser+2016}.
{\sc spexxy} is a full spectrum fitting framework written in {\sc python} with the purpose to derive the main stellar parameters from an observed spectrum by fitting a grid of model spectra.
For this work we utilised the {\sc phoenix} library of synthetic spectra\footnote{\url{http://phoenix.astro.physik.uni-goettingen.de}} \citep{husser+2013}, computed for Solar-scaled chemical composition and convolved with the MUSE line spread function, kindly provided by Tim-Oliver Husser.
With this setup we derived the radial velocities, metallicities, and effective temperatures for all of our good quality sources, while keeping the stellar surface gravities fixed from expectations according to the Dartmouth set of isochrones\\footnote{\url{http://stellar.dartmouth.edu}} \citep{dotter+2008}.  The median radial velocity uncertainty for individual stars is $2.8$\,\kms and the median \feh~uncertainty is $0.08$\,dex.
The radial velocity estimates are corrected for perspective effects using the equations from \citet{vandeven+2006}, which are within $\pm0.2$\,\kms for our field of view.
We took M54's bulk proper motion from \citet[][$\rm PM_{RA} = -2.683\pm0.025\,mas/yr, PM_{DEC} = -1.385\pm0.025$\,mas/yr]{vasiliev+baumgardt2021}.

We note that there is a significant difference in the setup for deriving stellar metallicities used in this work and in Paper~1 and hence there are inevitable systematic differences.
In this work we use a different fitting software and adopt a different spectral library with significantly larger wavelength coverage.
We also mask out the Na doublet at $5892$\,\AA, which was included in the metallicity fits in Paper~I.
Most notably, on the metal-poor end of the metallicity distribution, we measure consistently higher metallicities with respect to Paper~I in the order of $0.2$\,dex.
Although both methodologies relied on Solar-scaled models, the current models extend significantly further to the red and include the region of the Ca triplet, which can be heavily affected by non-Solar $\rm[\alpha/Fe]$ abundances.
Thus the systematically higher metallicities derived here are not surprising, as the metal-poor M\,54 stars are expected to be $\alpha$-enhanced.
At the metal-rich end of the metallicity distribution both methods yield consistent results.

The complete dataset used in this work (Appendix \ref{sec:app2}, Table \ref{tab:dataset}) is available online.

\section{Discrete population-dynamical models}\label{sec:dyn}

In the dynamical analysis of M\,54 we utilise the Jeans Anisotropic Multi-Gaussian Expansion (JAM) method initially developed by \citet{cappellari2008}.
The Jeans equations \citep{jeans1915} are derived from the steady-state collisionless Boltzmann equation and describe statistically the motion of a large collection of stars in a gravitational potential.
Under the assumption of axisymmetry, there are two equations, which link the functional form of the gravitational potential to the stellar density distribution of the system and the velocity moments of the stars.

In our work we use the {\sc python} version of the axisymmetric JAM code, written by M. Cappellari\footnote{\url{https://www-astro.physics.ox.ac.uk/~cappellari/software/\#jam}} \citep{cappellari2008,cappellari2012}.
According to Paper~II, M54's rotation axis, which is also the symmetry axis in our models, is coincidentally oriented along the sky meridian and thus vertically on the MUSE field of view.
We also assume that we see M\,54 edge-on, based on the 3D velocity analysis with Gaia proper motions published in Paper~II.

We build multi-population models with discrete posterior distribution functions \citep{watkins+2013}, following an approach similar to \citet{zhu+2016, kamann+2020}, that describe the distinct kinematic and morphological signatures of multiple stellar populations self-consistently.
The models are fully probabilistic and follow a Bayesian framework, i.e., we estimate the probability of each star from our observational sample to belong to a stellar population ($k$), based on its coordinates, radial velocity, age, and metallicity, given the respective uncertainties of these quantities.
There are three ingredients that describe the different stellar populations $k$ and constitute our dynamical models, which we describe as probability density functions (PDF):
\begin{enumerate}[i)]
\item $P_{pop}^k$ - a stellar population PDF, based on the inferred expected metallicity (and age) distributions of each population. In the two-population models, $P_{pop}^k$ is Gaussian, characterised by a mean metallicity and a metallicity spread. In the three-population model, $P_{pop}^k$ is a multi-variate Gaussian, characterised by a mean metallicity, a metallicity spread, and mean F606W \& F814W magnitudes per star, given an isochrone model, and the respective photometric colour spread (see Sect. \ref{sec:m2} for more details).
\item $P_{spa}^k$ - the spatial distribution or surface luminosity density of each population, which is inferred from the solution of the Jeans equations under the condition that the total surface luminosity density of all populations ($\sum\limits_k P_{spa}^k + $ foreground) is equal to the observed one.
\item $P_{dyn}^k$ - each population's velocity moments distribution - also inferred from the solution of the Jeans equations. Assuming a certain shape of the gravitational potential and the population's angular momentum ($\kappa^k$) and anisotropy ($\beta_z^k$), the JAM code predicts the first ($\overline{v}$) and second ($\overline{v^2}$) velocity moments at the position of each star in the sample. We compute the probability of each star's radial velocity to be drawn from a Gaussian with the first line-of-sight velocity moment as the mean and a variance $\sigma^2 = \overline{v^2} - \overline{v}^2$. The anisotropy is defined as $\beta_z = 1 - \frac{\overline{v_z^2}}{\overline{v_R^2}}$, where $\frac{\overline{v_z^2}}{\overline{v_R^2}}$ is the ratio of the second velocity moments in the vertical and radial direction in cylindrical coordinates. The angular momentum parameter $\kappa$ is a dimensionless quantity that provides a direct measure of the population's rotation, defined as $\kappa=\frac{\overline{v_{\phi}}}{(\overline{v_{\phi}^2} - \overline{v_R^2})^{1/2}}$, where $\overline{v_{\phi}}$ and $\overline{v_{\phi}^2}$ are the first and second velocity moments in the tangential direction. When $\kappa = \pm1$ and $\beta_z = 0$, the system reduces to an isotropic rotator and when $\kappa=0$, there is no net angular momentum. Negative $\beta_z$ values indicate tangential anisotropy, while positive $\beta_z$ values indicate radial anisotropy \citep{cappellari2008}. Although both the anisotropy and rotation parameters can vary with radius in GCs \citep{bianchini+2013,bianchini+2017b}, in this work we assume that they are radially constant, but independent for each population.
\end{enumerate}

\begin{figure}
	\includegraphics[width=\columnwidth]{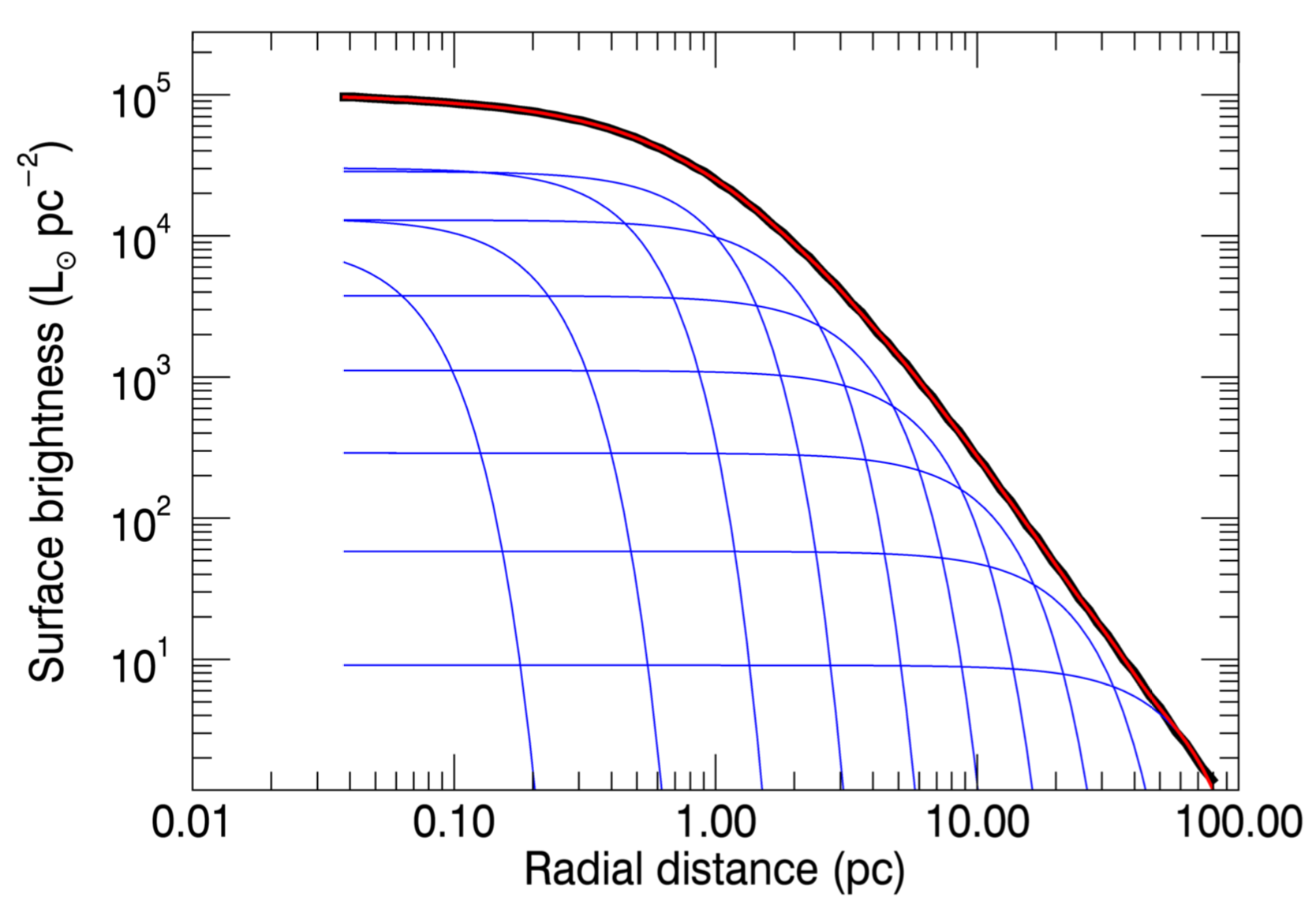}
    \caption{MGE fit (solid red line) to the observed surface brightness profile of M\,54 from \citet[][thick black line]{noyola+gebhardt2006}. The individual Gaussian components are plotted with solid blue lines.}
    \label{fig:m1_mge}
\end{figure}

The JAM code uses Multi-Gaussian Expansion (MGE) models to describe the surface luminosity density and the gravitational potential \citep{emsellem+1994}.
We use the V-band surface brightness profile of \citet{noyola+gebhardt2006} as representative of the system and fit it with a 10-component MGE model (Figure~\ref{fig:m1_mge}), using the MGE implementation and software of \citet{cappellari2002}.
According to Harris' catalogue (2010 version) the integral magnitude of M\,54 is $\rm M_V=-9.98$\,mag, which corresponds to an integral luminosity of $\rm L_V=0.85\times10^6$\,\Lsun.
We scale the MGE to match this figure.

We also include in all models an additional population component with a flat surface density, that represents the foreground stellar distribution. 
Its radial velocity moments are derived from the Besan\c{c}on model of the Galaxy \citep{robin+2003} at the position of M\,54, that is integrated along the line-of-sight. ($\langle V\rangle=40$\,\kms, $\sigma_V = 90$\,\kms).
The foreground population is expressed with a single free parameter ($\epsilon$), as a fraction of the central surface brightness of M\,54, as in \citet{watkins+2013, zhu+2016}.
Hence we can write the spatial PDFs of the different model components as:
\begin{equation}
P_{spa}^{k} = S_k / \sum_k (S_k + \epsilon C_k)
\end{equation}
and
\begin{equation}
P_{spa}^{foreground} = 1 - \sum_k P_{spa}^{k} ~,
\end{equation}
where $S_k$ is the surface brightness profile and $C_k$ - the central surface brightness of the various modelled populations ($k$).

The posterior probability of a star ($i$) to belong to either of the dynamical components or to the foreground component of the model ($k$) is then given by the joint probability of the above specified three probability distributions and we maximise the $\log$ posterior function of the entire sample:
\begin{equation}
\ln P = \sum_i \ln \sum_k P_{pop,i}^k P_{spa,i}^k P_{dyn,i}^k + \ln P_{pri}~,
\end{equation}
where $P_{pri}$ contains the parameter priors (see below).
We use the {\sc emcee} affine invariant Markov Chain Monte Carlo (MCMC) algorithm \citep{goodman+weare2010} in {\sc python} \citep{foreman-mackey+2013}.
We run our models on a CPU cluster engaging 96 cores and use 192 walkers and 6000 steps in the MCMC, which we confirmed to be enough for the fit to converge.

We explore dynamical models with two and three population components.
In the two-population model we use only the measured metallicity of each star in the sample as a population tag, while in the three-population model we also include the F606W and F814W HST magnitudes and fit for the population ages using isochrones.
The former model is simpler in construction and it aims to describe the interplay between the metal-poor and metal-rich stars in the M\,54 system, but ignores the Sgr's field population, which is at large mixed with the metal-rich nuclear component, but has different spatial and kinematic properties.
The three-population model, on the other hand, aims to provide a more complete physical picture of this dynamically complex system, exploring the interplay of several population-dynamical components simultaneously, but requires more prior assumptions to converge to a physical solution.

\subsection{A two-population model}\label{sec:m1}

At first we consider a population-dynamical model with two distinct stellar populations, separated by metallicity.
We assume that M\,54 consists of a metal-poor (MP) and a metal-rich (MR) component and we fit for their mean metallicities and intrinsic metallicity spreads, which gives 4 free parameters.

To model the spatial probability distributions of the two populations, we allow them to have decoupled surface brightness profiles, by setting each population to contribute an independent fraction to the luminosity of each of the 10 MGE components, that define the surface brightness of the entire nuclear system.
This appraoch ensures that the sum of the two decoupled surface brightness profiles is equal to the observed one from \citet{noyola+gebhardt2006}.
In principle, the JAM models can also take a flattening value ($q = \frac{b}{a}$, where $a$ and $b$ are the major and minor axes, respectively) for each MGE component, but here we adopt a radially-constant flattening for each population, and assume that the two flattening parameters are independent.
This leads to 12 additional free model parameters - one fraction for each MGE component and two flattening values - $\rm q_{MP}$ \& $\rm q_{MR}$ for the metal-poor and metal-rich populations, respectively.

For the gravitational potential we first assume that mass follows light and introduce a constant mass-to-light ratio ($\rm M/L_V \equiv \Upsilon_V$) for each of the two populations as free parameters.
We impose weak Gaussian priors with relatively large variance on these quantities ($\rm\langle\Upsilon_V^{MP}\rangle = 2.2\,M_{\odot}/L_{\odot}$, $\rm\sigma_{\Upsilon}^{MP}=0.5\,M_{\odot}/L_{\odot}$ and $\rm\langle\Upsilon_V^{MR}\rangle = 1.1\,M_{\odot}/L_{\odot}$, $\rm\sigma_{\Upsilon}^{MR}=0.5\,M_{\odot}/L_{\odot}$) so that they are loosely consistent with expectations from stellar population synthesis models.
According to the BaSTI\footnote{\url{http://albione.oa-teramo.inaf.it}} stellar population models \citep{percival+2009}, an old MP population (age = $11.5$\,Gyr, $\feh=-1.27$) is expected to have $\rm\Upsilon_V = 2.2\,M_{\odot}/L_{\odot}$ and a young MR population (age = $2.5$\,Gyr, $\feh=-0.25$) is expected to have $\rm\Upsilon_V = 1.1\,M_{\odot}/L_{\odot}$.
The total mass profile of M\,54 can then be computed in the following way:
\begin{equation}
\rm M(r) = \Upsilon_{MR}\,L_{MR}(r) + \Upsilon_{MP}\,L_{MP}(r)
\end{equation}
We explore a gravitational potential with radially variable $\Upsilon(r)$ later in Sect. \ref{sec:m1.2}.

In addition, we set as free parameters dimensionless quantities that describe each population's angular momentum ($\kappa_{MR}, \kappa_{MP}$) and velocity anisotropy ($\beta_{z,MR}, \beta_{z,MP}$).
We also set the systemic velocity (RV) of the system as a free parameter in this model ($\langle V\rangle$).
An additional parameter controls the foreground fraction of stars ($\epsilon$).
This leaves us with a total of 24 free model parameters.

\begin{figure*}
	\includegraphics[width=18cm]{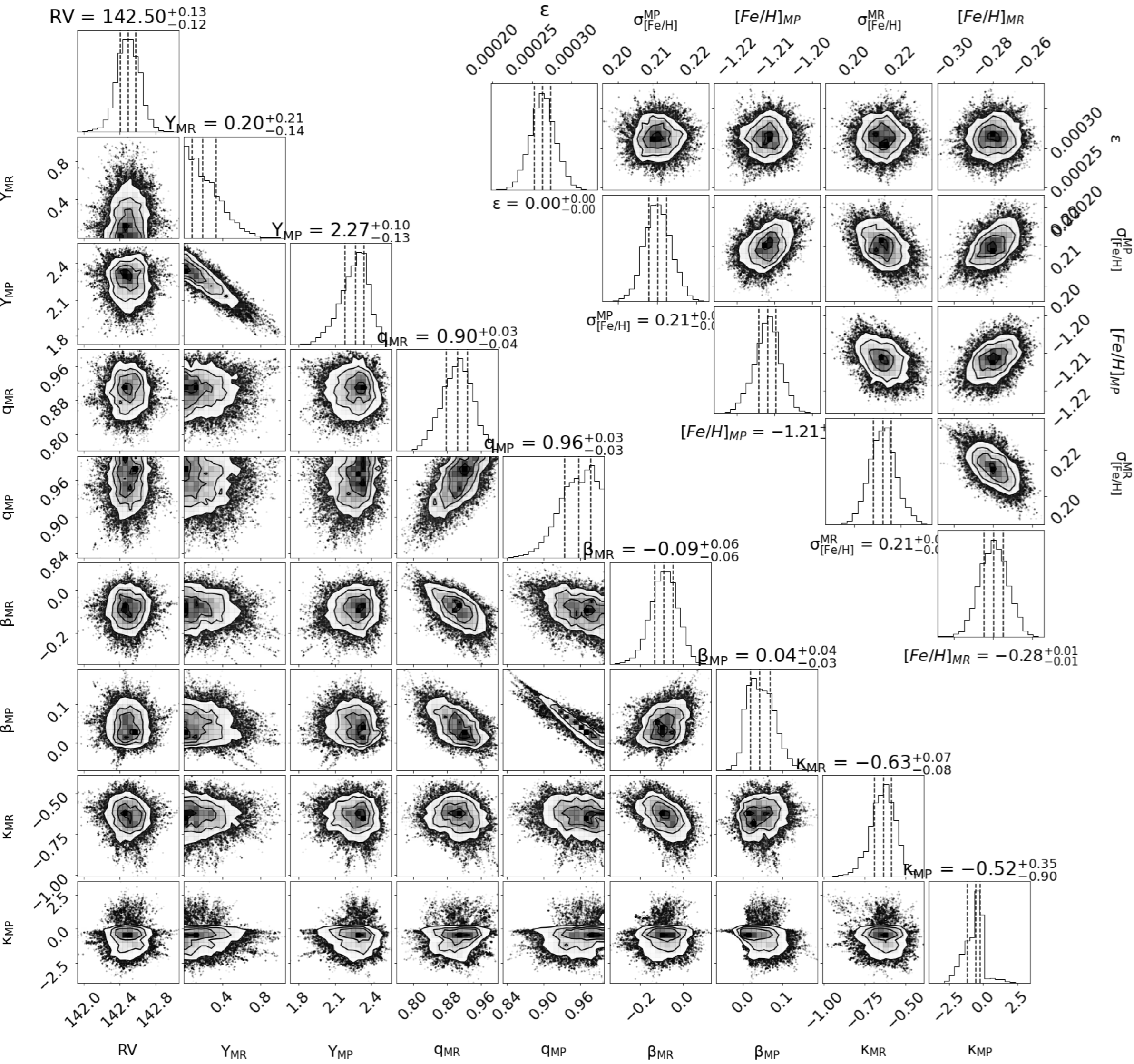}
    \caption{MCMC corner plot showing the posterior distributions of the 2-population dynamical model parameters with two constant M/L ratios (Sect. \ref{sec:m1}). The 10 light fractions of the surface brightness MGE are omitted.}
    \label{fig:m1_corner}
\end{figure*}

A corner plot showing the posterior distributions of the 14 main parameters is presented in Figure~\ref{fig:m1_corner}.
Their values are summarised in Table \ref{tab:dyn_param}.
In Appendix \ref{sec:app1} (Figure \ref{fig:mge_frac}) we also show the covariances of the 10 additional parameters describing the MGE fractions contributing to the MP population (the MR population fractions are one minus those, respectively).
One could note from Figure \ref{fig:mge_frac} that not all MGE components are relevant for separating the MP and MR populations' surface brightness distributions, however their combined effect, marginalised over the posterior, leads to a reliable and reproducible solution.

\begin{table}
	\centering
	\caption{Dynamical model best fit parameters - median and standard deviations from the MCMC posterior distribution.}
	\resizebox{9.5cm}{!}{
	\label{tab:dyn_param}
	\begin{tabular}{cccc}
		\hline
         & Sect. \ref{sec:m1} & Sect. \ref{sec:m1.2} & Sect. \ref{sec:m2}  \\
		\hline   
$\rm\langle V_0\rangle$ [\kms] & $142.50\pm0.13$  & $142.46\pm0.13$ & -- \\
$\rm\Upsilon_{MR} [M_{\odot}/L_{\odot,V}]$ &  $\rm 0.20\pm0.18$ & -- & -- \\
$\rm\Upsilon_{MP} [M_{\odot}/L_{\odot,V}]$ &  $\rm 2.27\pm0.11$ & -- & -- \\
$\rm\Upsilon_0 [M_{\odot}/L_{\odot,V}]$ &  -- & $\rm 5.42\pm2.62$ & $\rm 5.62\pm2.35$ \\
$\rm\Upsilon_t [M_{\odot}/L_{\odot,V}]$ &  -- & $\rm 1.10\pm0.25$ & $\rm 1.00\pm0.15$\\
$\rm\Upsilon_{\infty} [M_{\odot}/L_{\odot,V}]$ &  -- & $\rm 2.89\pm0.29$ & $\rm 3.15\pm0.36$\\
$\rm r_t [\arcsec]$ &  -- & $3.98\pm1.25$ & $4.38\pm0.80$ \\
$\rm q_{YMR}$ & $0.90\pm0.04$ & $0.91\pm0.03$ & $0.84\pm0.03$ \\
$\rm q_{IMR}$ & -- & -- & $0.88\pm0.06$ \\
$\rm q_{OMP}$ & $0.96\pm0.03$ & $0.97\pm0.03$  & $0.95\pm0.04$ \\
$\rm \beta_z^{YMR}$ & $-0.09\pm0.06$ & $-0.08\pm0.06$ & $0.00\pm0.08$ \\
$\rm \beta_z^{IMR}$ & -- & -- & $0.09\pm0.14$ \\
$\rm \beta_z^{OMP}$ & $0.04\pm0.04$ & $0.04\pm0.03$ & $0.08\pm0.04$ \\
$\rm \kappa_{YMR}$ & $-0.63\pm0.08$ & $-0.65\pm0.08$ & $-0.98\pm0.17$ \\
$\rm \kappa_{IMR}$ & -- & -- & $-0.06\pm0.34$ \\
$\rm \kappa_{OMP}$ & $-0.52\pm0.53$ & $-0.31\pm0.52$ & $0.07\pm1.05$ \\
$\rm\langle[Fe/H]\rangle_{YMR}$ & $-0.28\pm0.007$ & $-0.28\pm0.007$ & $-0.21\pm0.004$ \\
$\rm\langle[Fe/H]\rangle_{IMR}$ & -- & -- & $-0.47\pm0.006$ \\
$\rm\langle[Fe/H]\rangle_{OMP}$ & $-1.21\pm0.003$  & $-1.21\pm0.003$ & $-1.21\pm0.004$ \\
$\rm\sigma_{[Fe/H]}^{YMR}$ & $0.21\pm0.006$ & $0.21\pm0.006$ & $0.17\pm0.008$ \\
$\rm\sigma_{[Fe/H]}^{IMR}$ & -- & -- & $0.26\pm0.02$ \\
$\rm\sigma_{[Fe/H]}^{OMP}$ & $0.21\pm0.003$ & $0.21\pm0.003$ & $0.24\pm0.005$ \\
$\rm\langle age\rangle_{YMR} [Gyr]$ & -- & -- & $2.53\pm0.003$ \\
$\rm\langle age\rangle_{IMR} [Gyr]$ & -- & -- & $5.37\pm0.06$ \\
$\rm\langle age\rangle_{OMP} [Gyr]$ & -- & -- & $14.14\pm0.13$ \\
$\rm\sigma_{V-I}^{YMR} [mag]$ & -- & -- & $0.05\pm0.01$ \\
$\rm\sigma_{V-I}^{IMR} [mag]$ & -- & -- & $0.0005\pm0.0004$ \\
$\rm\sigma_{V-I}^{OMP} [mag]$ & -- & -- & $0.0003\pm0.0003$ \\
$\rm E(B-V) [mag]$ & -- & -- & $0.11\pm0.0005$ \\
$\rm\epsilon [\times10^{-4}]$ & $2.6\pm0.15$ & $2.7\pm0.15$ & $3.1\pm0.15$ \\
		\hline
	\end{tabular}
	}
\end{table}

We find dynamical $\rm \Upsilon_V^{MP} = 2.3\pm0.1\,M_{\odot}/L_{\odot,V}$ and $\rm \Upsilon_V^{MR} = 0.2\pm0.2\,M_{\odot}/L_{\odot,V}$.
$\rm \Upsilon_V^{MP}$ is consistent with expectations from synthesis models of old ($12$\,Gyr) MP stellar populations at this metallicity according to the BaSTI stellar population synthesis tool.
$\rm \Upsilon_V$ for the metal-rich stars of M\,54 on the other hand is significantly lower than expectations for $1.5 - 2.0$\,Gyr MR ($\feh=-0.25$\,dex) populations synthesis models and consistent with zero.
The model tends to attribute all mass to only one of the two populations.

With this perhaps overly simplistic assumption of constant $\rm \Upsilon$ ratios for the two stellar populations, our model predicts a total mass of the system of $1.26\pm0.03\times10^6$\,\Msun~at the tidal radius.
The mass of the MP population is $96\%$ of the total, while it accounts for only $64\%$ of the luminosity.
In result we estimate a global $\rm \Upsilon_V = 1.51\pm0.04$\,\MLsun for the entire system, which is in line with expectations for a mixed system consisting of both old and young stellar populations.

We also ran a model with a single constant mass-to-light ratio describing the gravitational potential for the entire system. 
Not surprisingly, we find a $\rm \Upsilon_V = 1.52\pm0.03\,M_{\odot}/L_{\odot,V}$, which is very similar to the global value found above.
We note that in this test we did not impose any prior on $\rm \Upsilon$.

\begin{figure}
	\includegraphics[width=\columnwidth]{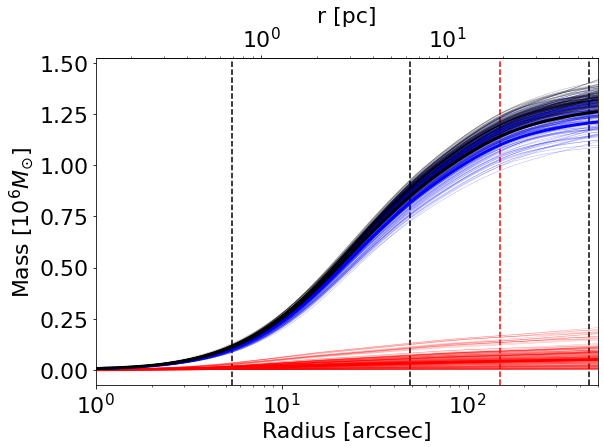}
    \caption{Cumulative masses of the two populations (blue - MP; red - MR) and the entire system (black), according to the model described in Sect. \ref{sec:m1}. The three black vertical dashed lines on both panels correspond to M\,54's core, half-light, and tidal radii \citep[GC catalogue of][2010 edition]{harris1996}, while the red dashed line indicates the end of the MUSE mosaic field of view.}
    \label{fig:m1_pop_mass}
\end{figure}

When we refer to the total dynamical mass in this work, we consider the mass locked within the tidal radius of the system, as given in the Harris catalogue of globular clusters \citep[see Table \ref{tab:param};][2010 edition]{harris1996}, which corresponds to a radial distance of $\sim77$\,pc.
However, these values need to be taken cautiously for two main reasons.
Firstly, our kinematic data does not reach that far out in the system and therefore, the quoted quantities are extrapolated predictions of the model.
Our MUSE mosaic field of view contains $\sim85\%$ of the total luminosity of the nuclear system.
%Secondly, nuclear clusters generally do not have a defined tidal radius, unless they are fully tidally stripped.
Secondly, despite feeling the Milky Way tidal field, M\.54's nucleus is still embedded in the densest central region of the disrupting Sgr dSph and hence its tidal radius is not well defined.
The tidal radius that we quote comes simply from the best fit \citet{king1962} profile to the surface brightness profile of the system by \citet{mclaughlin+vandermarel2005}.
We stress that the surface brightness profile that we use in our dynamical models (approximated with a MGE) is not truncated at this radius.
In this respect it is much more meaningful to look at the cumulative mass profile that we obtain with our dynamical models and present in Figure \ref{fig:m1_pop_mass}.
We present the estimated masses of the two populations and the entire system at several radial distances in Table \ref{tab:m1_masses}.

We conclude that with this choice of gravitational potential parametrisation, the population-dynamical model is prone to attributing all of the mass to a single component and hence we next explore dynamical models with a single radially varying $\Upsilon$.
To avoid repetition, we discuss the other results of the 2-population dynamical model and their implications in the next section, as they appear practically unaffected by the choice of the gravitational potential parametrisation.

\begin{table*}
	\centering
	\caption{Cumulative dynamical mass estimates and $\rm\Upsilon_V$ of M\,54's populations according to the model presented in Section \ref{sec:m1}. The various radii are defined in Table \ref{tab:param}.}
	%\resizebox{\textwidth}{!}{
	\label{tab:m1_masses}
	\begin{tabular}{ccccccccc }
		\hline
        & MP & MR & Total & \% MP & \% MR & $\rm \Upsilon_{MP}$ & $\rm \Upsilon_{MR}$ & $\rm \Upsilon_{Tot}$ \\
        &  [$10^6$\,\Msun]  & [$10^6$\,\Msun]  &  [$10^6$\,\Msun]  &      &      & [\MLsun]     &  [\MLsun]   & [\MLsun] \\	     
		\hline   
$\rm r_c$ &  $0.10\pm0.01$ & $0.01\pm0.01$ & $0.11\pm0.01$ & 92 & 8 & $2.27\pm0.08$ & $0.20\pm0.17$ & $1.20\pm0.05$ \\
$\rm r_h$ &  $0.82\pm0.02$ & $0.03\pm0.03$ & $0.85\pm0.02$ & 96 & 4 & $2.27\pm0.04$ & $0.20\pm0.19$ & $1.53\pm0.04$ \\
$\rm r_{fov}$ &  $1.10\pm0.05$ & $0.04\pm0.04$ & $1.14\pm0.03$ & 96 & 4 & $2.27\pm0.04$ & $0.20\pm0.19$ & $1.55\pm0.04$ \\
$\rm r_t$ &  $1.21\pm0.06$ & $0.05\pm0.05$ & $1.26\pm0.03$ & 96 & 4 & $2.27\pm0.06$ & $0.20\pm0.19$ & $1.51\pm0.04$ \\
		\hline
	\end{tabular}
	%}
\end{table*}

\subsection{Radially varying mass-to-light ratio}\label{sec:m1.2}

\begin{figure*}
	\includegraphics[width=18cm]{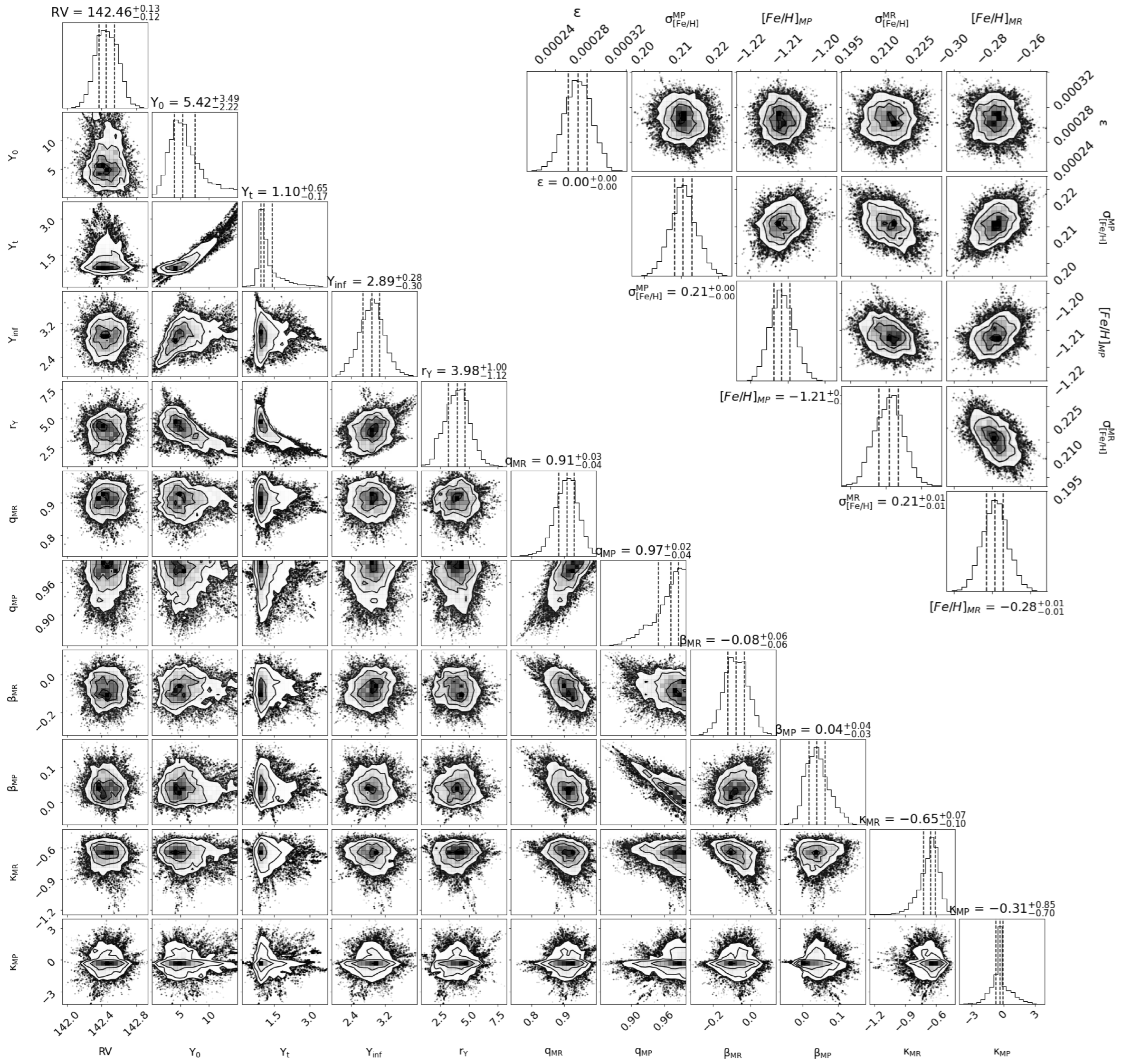}
    \caption{MCMC corner plot showing the posterior distributions of the 2-population dynamical model parameters with a radially varying M/L ratio (Sect. \ref{sec:m1.2}). The 10 light fractions of the surface brightness MGE are omitted.}
    \label{fig:m12_corner}
\end{figure*}

The population-dynamical model discussed in the previous section has one significant drawback, which is the tendency to attribute the entire mass budget to only one of the model components.
Our solution is to model the gravitational potential globally for the entire system, but allow for more degrees of freedom when fitting for it, by introducing a radially varying universal mass to light ratio - $\rm \Upsilon(r)$.

We adopted a similar technique to \citet{kamann+2020} to model the gravitational potential.
Essentially, we assigned an individual $\rm \Upsilon$ to each MGE component in the surface brightness profile of M\,54, according to a parametrisation given by Equation 5 in \citet{kamann+2020}:
\begin{equation}
\label{eq:ml}
{\rm \Upsilon(r) = \frac{\Upsilon_0\left(1-\frac{r}{r_{\Upsilon}}\right) + 2\Upsilon_t\left(\frac{r}{r_{\Upsilon}}\right) - \Upsilon_{\infty}\frac{r}{r_{\Upsilon}}\left(1-\frac{r}{r_{\Upsilon}}\right)}{1+\left(\frac{r}{r_{\Upsilon}}\right)^2} }
\end{equation}
where $\rm \Upsilon_0$ is the M/L ratio in the centre of the system, $\rm \Upsilon_{\infty}$ - at infinity, and $\rm \Upsilon_t$ is a transition value at a characteristic radius - $r_{\Upsilon}$.
The $\rm \Upsilon$ value corresponding to each MGE component is calculated at a radius of $1\,\sigma$ of that component, where $\sigma$ is the standard deviation width of each Gaussian component, comprising the MGE.
Hence, the resulting $\rm M/L_V$ profile from dividing the mass $\rm M(r)$ and luminosity $\rm L_V(r)$ profiles from the MGE components is not exactly the same as $\rm\Upsilon(r)$ in Equation \ref{eq:ml}.
The functional form of the radially varying $\rm \Upsilon$ is motivated by GC dynamical models \citep{denBrok+2014a, bianchini+2017} and $N$-body simulations \citep{baumgardt2017}, which show that the mass-to-light ratio radial profile of GCs is cup-shaped with a peak in the centre, a gradual decrease towards the half-mass radius, followed by another increase in the outer regions.
Equation \ref{eq:ml} mimics this behaviour.
These three $\rm \Upsilon$ figures and the characteristic radius are free parameters in our model.
\citet{baumgardt2017} note that for the majority of Galactic GCs the characteristic radius ($r_{\Upsilon}$) is at about a tenth of the half-mass radius ($\rm r_h \simeq 50\arcsec$ or $6.8$\,pc), so we introduce a Gaussian prior with this condition ($\langle r_{\Upsilon}\rangle = 5\arcsec$, $\sigma_{r_{\Upsilon}}=1\arcsec$), otherwise $\rm \Upsilon_t$ and $r_{\Upsilon}$ are degenerate.
Furthermore, as our data is spatially limited and does not extend to the tidal radius of M\,54, we do not have a good constraint on $\rm \Upsilon_{\infty}$, so we also set a Gaussian prior on this parameter with a mean of $\rm \Upsilon_{\infty} = 2\,$\MLsun~and a standard deviation $0.5$\,\MLsun, which is typical for globular clusters and ultra-compact dwarfs old stellar populations \citep{voggel+2019}.

We kept all other model parameters the same as in Sect. \ref{sec:m1}, which results into a population-dynamical model with 26 free model parameters.
A corner plot showing their covariances is presented in Figure \ref{fig:m12_corner} and their values are summarised in Table \ref{tab:dyn_param}.

\begin{figure}
	\includegraphics[width=\columnwidth]{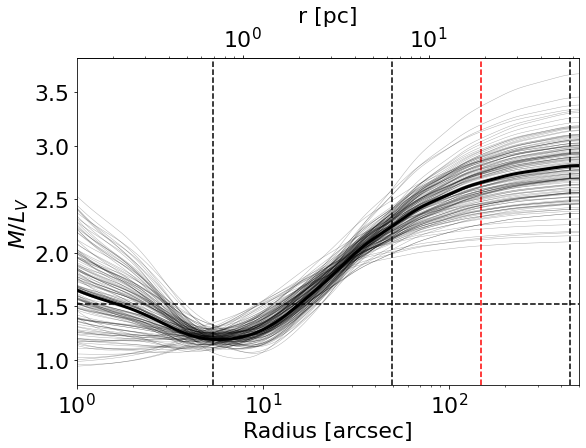}
	\includegraphics[width=\columnwidth]{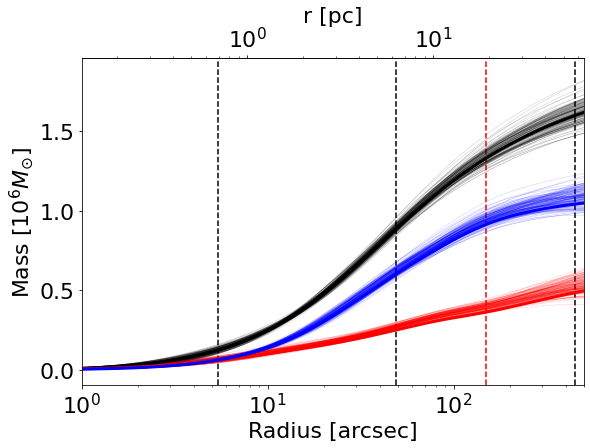}
    \caption{Dynamical results according to the model described in Sect. \ref{sec:m1.2}. {\it Top:} M\,54 radial $\rm \Upsilon_V$ profile, resulting from dividing the best fit mass and luminosity MGE models of the system. The median $\rm \Upsilon_V$ profile is indicated with a thick curve, while the thin curves represent random draws from the posterior distribution. The horizontal dashed line corresponds to the best fit constant $\rm \Upsilon_V$. {\it Bottom: } Cumulative masses of the two populations (blue - MP; red - MR) and the entire system (black). The three black vertical dashed lines on both panels correspond to M\,54's core, half-light, and tidal radii \citep[GC catalogue of][2010 edition]{harris1996}, while the red dashed line indicates the end of the MUSE mosaic field of view.}
    \label{fig:m1_ml_var}
\end{figure}

Figure~\ref{fig:m1_ml_var} shows the resulting radial $\rm \Upsilon_V$ ratio profile and the estimated cumulative mass distribution of M\,54.
The best fit model indeed prefers a slightly higher $\rm \Upsilon$ in the innermost region of the system with a minimum at the core radius, followed by a gradual increase outwards.
It is assumed that the mass segregation of stellar remnants is the reason for the upturn in M/L at the centre \citep{bianchini+2017}, while $\Upsilon$ is high in the outer parts due to the over-density of lower mass stars.
In the more complicated case of M\,54, the presence of the more centrally concentrated MR, significantly younger population (expected lower $\rm \Upsilon_V$) works to mitigate the effect of mass segregation.
At the same time, the entire system is likely embedded in a dark matter halo \citep{carlberg+2022}, not surprising for a dwarf galaxy nucleus, which further increases the mass-to-light ratio in the outer regions.
%It is interesting to note that $\rm \Upsilon_V(r)$ in the outer parts (where the MP stars dominate the gravitational potential) is equal to $\rm \Upsilon_V^{MP}$ for the MP stars in the dynamical model discussed in the previous section.
%In the inner region, however, where both MP and MR stars contribute substantially to M\,54's gravitational potential, $\rm \Upsilon_V(r)$ is lower on average but larger than $\rm \Upsilon_V^{MR}$ we found for the MR population in Sect. \ref{sec:m1}.

Due to the change in the shape of the potential, compared to the dynamical model described in Sect. \ref{sec:m1}, we find a higher mass for the M\,54 system ($\rm M_{tot} = 1.58\pm0.07\times10^6$\,\Msun~at the tidal radius), where $67\%$ of the mass and luminosity are attributed to the MP population.
We note that since in this dynamical model we use a common M/L ratio for both populations, the mass contribution of each population mostly follows its luminosity contribution.
Small differences arise due to the different spatial extent of the luminosity density distributions of the two populations.
We find a global $\rm \Upsilon_V = 1.92\pm0.08$\,\MLsun~for the entire system, using a 2-population dynamical model with a radially variable mass-to-light ratio.
Table \ref{tab:m1.2_masses} gives the mass estimates and $\rm\Upsilon_V$ of the two populations locked within several key radial points.

\begin{table*}
	\centering
	\caption{Cumulative dynamical mass estimates and $\rm\Upsilon_V$ of M\,54's populations according to the model presented in Sect. \ref{sec:m1.2}. The various radii are defined in Table \ref{tab:param}.}
	%\resizebox{\textwidth}{!}{
	\label{tab:m1.2_masses}
	\begin{tabular}{ccccccccc }
		\hline
        & MP & MR & Total & \% MP & \% MR & $\rm \Upsilon_{MP}$ & $\rm \Upsilon_{MR}$ & $\rm \Upsilon_{Tot}$ \\
        &  [$10^6$\,\Msun]  & [$10^6$\,\Msun]  &  [$10^6$\,\Msun]  &      &      & [\MLsun]     &  [\MLsun]   & [\MLsun] \\	     
		\hline   
$\rm r_c$ &  $0.06\pm0.01$ & $0.06\pm0.01$ & $0.12\pm0.01$ & 50 & 50 & $1.32\pm0.07$ & $1.36\pm0.12$ & $1.35\pm0.14$ \\
$\rm r_h$ &  $0.61\pm0.02$ & $0.26\pm0.01$ & $0.87\pm0.02$ & 70 & 30 & $1.65\pm0.03$ & $1.51\pm0.06$ & $1.59\pm0.04$ \\
$\rm r_{fov}$ &  $0.93\pm0.03$ & $0.38\pm0.02$ & $1.31\pm0.04$ & 71 & 29 & $1.85\pm0.03$ & $1.71\pm0.06$ & $1.81\pm0.06$ \\
$\rm r_t$ &  $1.06\pm0.04$ & $0.52\pm0.04$ & $1.58\pm0.07$ & 67 & 33 & $1.93\pm0.05$ & $1.90\pm0.11$ & $1.92\pm0.08$ \\
		\hline
	\end{tabular}
	%}
\end{table*}

We can also immediately see from Figures~\ref{fig:m1_corner} \& \ref{fig:m12_corner}  that the MP population is close to isotropic ($\beta_z = 0.04\pm0.04$) with considerably lower angular momentum than the MR one.
The latter also has a negative $\beta_z=-0.09\pm0.06$ parameter, which is indicative of a slight tangential anisotropy \citep{cappellari2008}.

\begin{figure*}
	\includegraphics[width=\columnwidth]{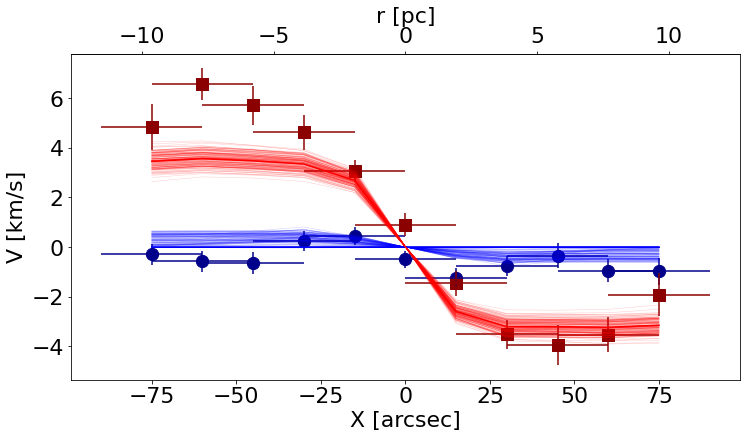}
	\includegraphics[width=\columnwidth]{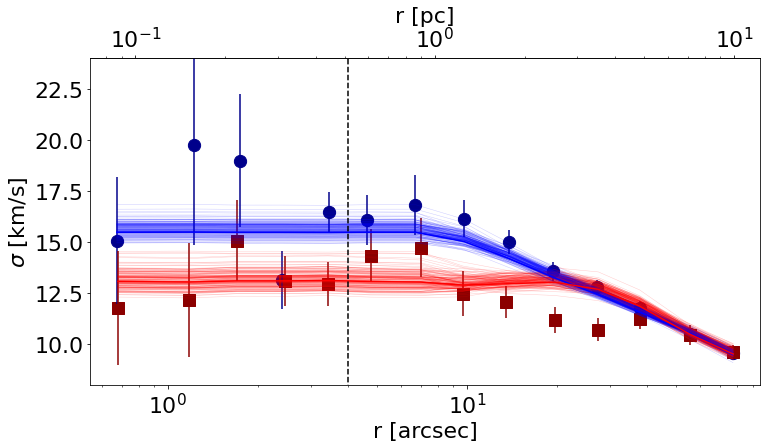}
    \caption{Predictions for the rotation ({\it left panel}) and the velocity dispersion ({\it right panel}) for the two populations drawn from the model posterior (MR - red lines, MP - blue lines) according to the dynamical model described in Sect. \ref{sec:m1.2}, compared to the observed binned profiles (red squares and blue dots with error bars, respectively), using stars with $>50\%$ probability to belong to either of the two populations. The horizontal error bars indicate the size of each bin. The vertical dashed line in the {\it right panel} shows the border between the NFM and WFM MUSE observations.}
    \label{fig:m1_kin}
\end{figure*}

The discrete Jeans model predicts the velocity moments at the position of each observed star and the parameter optimisation was performed on the discrete data.
To visually compare the model prediction to the observed kinematics, we assign a parent population to each of the observed stars based on a 50\% probability threshold, which is only relevant for presentation purposes and plotting.
Then we bin both the observations and the models in the same way to produce radial rotation and velocity dispersion profiles (Figure~\ref{fig:m1_kin}).
The choice of binning is purely for visualisation purposes and does not have any effect on the results and conclusions in this work.
%We use bins that are parallel to the rotation axis for the rotation profile and concentric rings for the velocity dispersion profile.
We obtain the mean velocity and intrinsic velocity dispersion for each bin in Figure \ref{fig:m1_kin} via a maximum likelihood fit to the kinematic distribution using a Gaussian model with unknown mean and variance.
The variance in the likelihood function includes the individual measurement uncertainties.
%We use a maximum likelihood approach when deriving the mean velocity and velocity dispersion values in the observed kinematic profiles to properly account for the uncertainties in the individual radial velocity measurements, assuming that the velocity distribution of stars in each bin is normal with unknown mean and variance. 
We also correct the observed velocity dispersion profile for the systemic rotation.
For the rotation profile we use overlapping linearly spaced vertical slices along the horizontal x-axis (Figure~\ref{fig:m1_kin}, left panel) and for the velocity dispersion we use logarithmically spaced concentric radial bins (Figure~\ref{fig:m1_kin}, right panel).
We perform multiple random draws from the model posterior to derive the model uncertainties.
Overall, the 2-component Jeans dynamical model describes the observed kinematics of the system at all radii very well.
It is evident that the MR population has a high degree of intrinsic rotation combined with a lower velocity dispersion, while the MP component has a significantly lower angular momentum and higher velocity dispersion.

\begin{figure*}
	\includegraphics[width=\columnwidth]{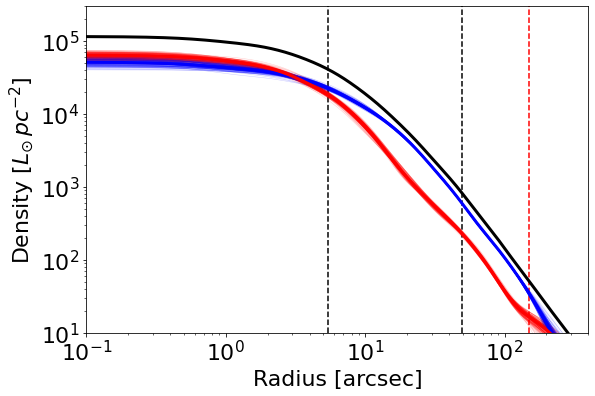}
	\includegraphics[width=\columnwidth]{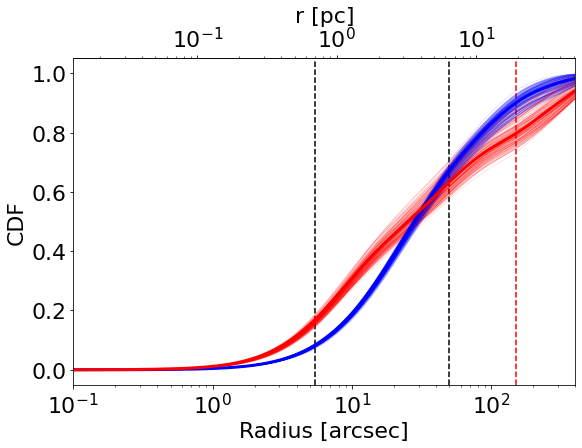}
    \caption{{\it Left panel:} Radial distribution functions of the two population-dynamical components of M\,54 as drawn from the Sect. \ref{sec:m1.2} model posterior (red lines for the MR and blue lines for the MP). The solid black line shows the combined surface brightness profile. {\it Right panel:} Radial cumulative distribution function of the two population-dynamical components of M\,54.}
    %{\it Bottom left panel:} Relative fractions of the two populations as a function of radius. {\it Bottom right panel:} Cumulative masses of the two populations and the entire system (black lines). The black dashed lines indicate the core, half-light, and tidal radii of the system and the red dashed line indicates the limit of the MUSE field of view.
    \label{fig:m1_rdf_cdf}
\end{figure*}

The radial distributions of the two populations, together with their cumulative distribution functions are presented in Figure~\ref{fig:m1_rdf_cdf}.
They are both very close to spherical with axis ratios $q_{\rm MR} = 0.90\pm0.04$ and $q_{\rm MP} = 0.96\pm0.03$ for the metal-rich and metal-poor populations, respectively.
Note that we reported larger flattening for both populations in Paper~I, using a direct elliptical Plummer model fit to their observed stellar number densities - $\rm q_{MR} = 0.69\pm0.10$ and $\rm q_{MP} = 0.84\pm0.06$.
The method used in Paper~I, however, is sensitive to incompleteness in the data, while here we infer the population flattenings purely from dynamical constraints.
It is important to note here that the modelled density profiles of the two populations are inferred solely from the Jeans equations and the observed stellar kinematics, and do not depend on the observed stars' spatial distribution, which is heavily biased due to the very uneven photometric depth reached by the different MUSE data sets.
In fact, there is no need that the observed kinematic tracers follow the actual density distribution of the stars in order to fit for their density profiles.
This makes our models insensitive to various sources of incompleteness in the observations and we can also predict the stellar density outside of the MUSE mosaic field of view, relying on the M\,54's surface brightness profile by \citet{noyola+gebhardt2006}, which extends much further out.
Figure~\ref{fig:m1_rdf_cdf} shows that the MP \& MR components have different spatial distributions.
The MR stars are generally more centrally concentrated than the MP stars, but their distribution flattens out significantly in the outer region.
We noted that the MR stars are more centrally concentrated than the MP stars already in Paper~I, but here we show that this is a prediction of the best fit dynamical model too.
The excess of MR stars in the outer region belong to the intermediate age metal-rich Sgr field stars, which we do not explicitly consider in this model and they appear mixed with the metal-rich M\,54 population.
In Section \ref{sec:m2} we introduce a three component dynamical model to alleviate this problem.

\begin{figure*}
	\includegraphics[width=\columnwidth]{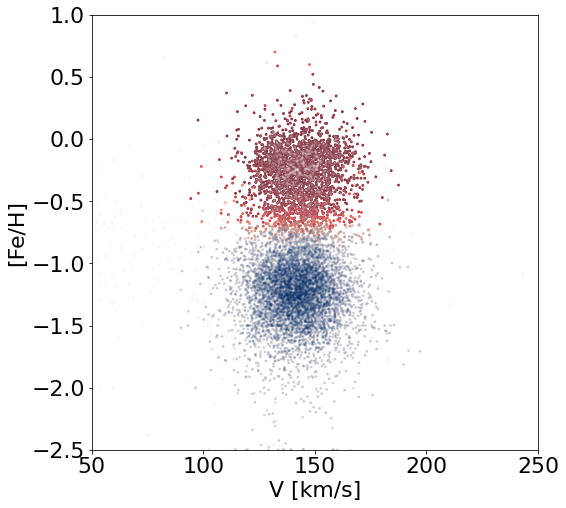}
	\includegraphics[width=8.1cm]{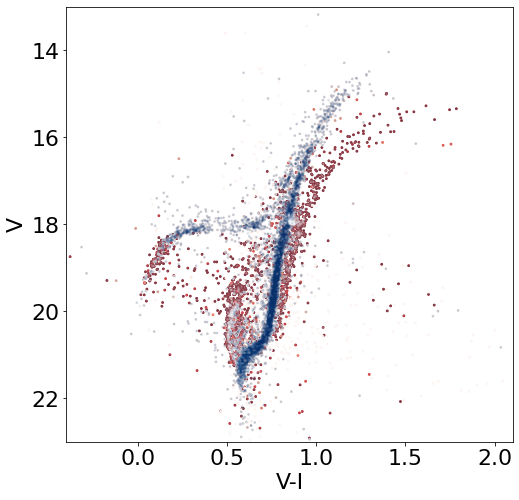}
    \caption{Metallicity vs. radial velocity plot of all stars in the sample ({\it left panel}) and a CMD ({\it right panel}), colour-coded by the probability of the star to belong to the MR population (shades of red) or the MP population (shades of blue), according to the dynamical model described in Sect. \ref{sec:m1.2}. The photometry is from \citet{siegel+2007}.}
    \label{fig:m1_pop}
\end{figure*}

In Figure~\ref{fig:m1_pop} we show the separation between the two dynamically distinct components using two population diagnostic plots.
The left panel of Figure~\ref{fig:m1_pop} shows M\,54's observed stellar sample in radial velocity vs. metallicity space, where individual stars are colour-coded with varying shades of blue and red for the MP and MR populations, respectively, depending on their membership probability.
The two populations with identical mean systemic velocities are well distinguishable on the \feh~axis.
We find a mean $\langle\feh_{\rm MR}\rangle = -0.28\pm0.01$\,dex and $\langle\feh_{\rm MP}\rangle = -1.21\pm0.01$\,dex with both populations having the same intrinsic metallicity spread of $0.21\pm0.01$\,dex.
The reported intrinsic metallicity spreads take into account the \feh~measurement uncertainties estimated with the {\sc spexxy} code, however there are indications that these uncertainties could be underestimated by $20\% - 30\%$ \citep{husser+2016}.
Still the measured intrinsic spreads are quite large and there are likely physical reasons for this.
The large \feh~spread in the MR population is likely due to inhomogeneous origin of the MR stars, which we explore in Sect. \ref{sec:m2}, while the large \feh~spread in the MP population could be explained by merging GCs with a spread in metallicities.

In the right panel of Figure~\ref{fig:m1_pop} we show the HST colour-magnitude diagram (CMD) of our sample, using photometry from \citet{siegel+2007} and the same population probability colour-coding.
As expected, the MP stars follow a narrow distribution on the CMD that is consistent with an old stellar population, similar to the majority of halo globular clusters (GC).
On the other hand, the MR stars are clearly younger and likely have a larger age spread.

The differences in kinematics, spatial distribution, and population properties are all well captured in a single self-consistent dynamical model, based on the Jeans equations.
The high angular momentum, lower dispersion, central concentration, and high metallicity of the MR stars all point to an {\em in situ} origin of this structure, where it originated in a rapidly rotating gaseous disk in the centre of the system.
The MR population appears marginally flatter than the MP population, although both are very close to spherical.
There are various mechanisms of disk heating, that could puff the MR stars to an almost spherical distribution.
For instance, the disk relaxes and redistributes its angular momentum to the stars in the MP population, becoming more spherical with time \citep[see Paper~II;][]{mastrobuono+hagai2013,mastrobuono+hagai2016}.
On the contrary, the MP population properties are fully consistent with GC origin, where one or more GCs arrived at the centre of Sgr through dynamical friction.

\subsection{A three-population model}\label{sec:m2}

\begin{figure*}
	\includegraphics[width=18cm]{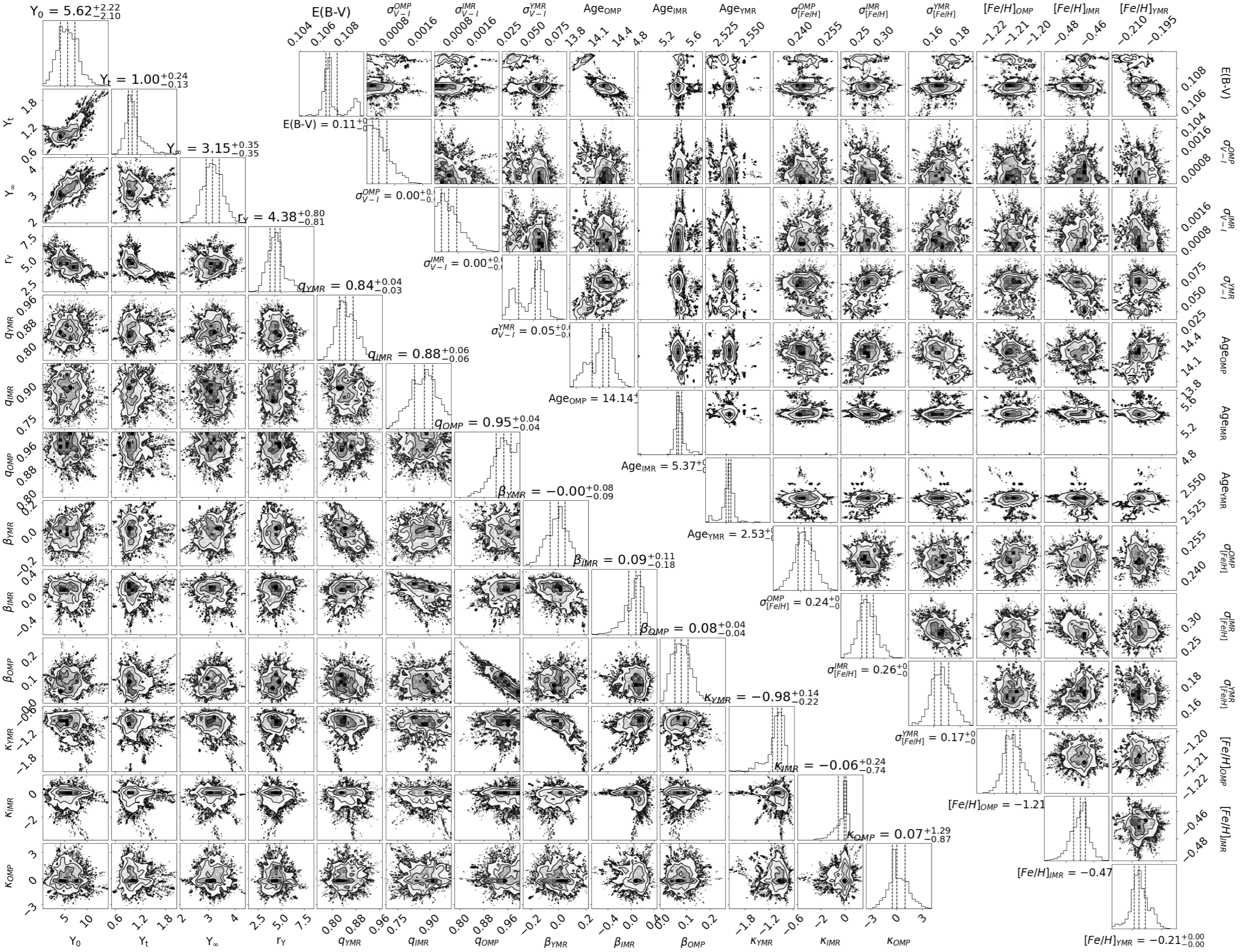}
    \caption{MCMC corner plot showing the posterior distributions of the 3-population-dynamical model parameters (Sect. \ref{sec:m2}). Covariances between parameters pertaining to the Jeans equations and isochrone fits are plotted separately.}
    \label{fig:m2_corner}
\end{figure*}

In Paper~I we identified the IMR population as representative of the Sgr field stellar content and showed that it has a more extended SFH and surface density distribution than the YMR and OMP populations.
It is also distinct with its flat velocity dispersion profile (Paper~II).

Here we present a 3-population dynamical model of M\,54, which includes simultaneously all three components identified in Papers~I\,\&\,II - the YMR, IMR, and OMP.
In order to better separate the different populations, we also consider their age difference and fit for the three populations mean ages.
To this aim we use the F606W and F814W magnitudes and their uncertainties from the \citet{siegel+2007} HST catalogue in addition to the measured stellar metallicities from the MUSE spectra and fit three isochrones simultaneously with the velocity moments from the Jeans equations to the three modelled populations.

We opted to work with the scaled-Solar Dartmouth set of isochrones and we developed an isochrone interpolator to this aim, that generates an isochrone for any custom combination of age and metallicity in the above mentioned photometric bands.
We tested isochrone grids with varying [$\alpha$/Fe], but found out that the scaled-Solar isochrones were a better match to the observed CMD, given the estimated stellar metallicities and fixed distance.
%We note that while it is expected that the OMP stars are more $\alpha$-enhanced than the YMR \& IMR stars, we measured the metallicities assuming a scaled-Solar spectral library, and thus the photometric effect of the $\alpha$-enhancement could be cancelled out by a reciprocal bias in the Solar-scaled metallicity estimates.
We note that while it is expected that the OMP stars are more $\alpha$-enhanced than the YMR \& IMR stars, we measured the metallicities assuming a scaled-Solar spectral library.
%In that respect, our model is consistent.
In addition we model the horizontal branch (HB) of the OMP population and the red clump (RC) of the YMR population using the Dartmouth Stellar Evolution Database HB/AGB Track Grids.
We chose the HB and RC models that best represent M\,54's CMD and kept them fixed, e.g. independent of the age and metallicity of the fitted isochrones.
We kept the assumed distance to M\,54, which is relevant for both the isochrone and velocity moments fits, fixed to $26.5$\,kpc, but set the line-of-sight reddening as a free parameter.
Over all, we fit for the mean metallicity and mean age of the YMR, IMR, and OMP populations, their corresponding intrinsic metallicity spreads, reddening, and we allow for a larger variance in the F606W$-$F814W colour in addition to the photometric uncertainties.
The latter could eventually be interpreted as coming from an intrinsic age spread of the individual populations, but such a connection is not straight-forward and we refrain from making it.
We assume a multi-dimensional Gaussian population likelihood for each observed star to belong to either of the three populations, based on its F606W, F814W magnitudes and metallicity, given an interpolated isochrone model defined by age, metallicity, and reddening.

We use the gravitational potential with a radially varying $\Upsilon$, as described in Section \ref{sec:m1.2} and adopt the same priors in the posterior function: $\langle r_{\Upsilon}\rangle = 5\arcsec$, $\sigma_{r_{\Upsilon}}=1\arcsec$ and $\rm \Upsilon_{\infty} = 2\,$\MLsun~with a standard deviation $0.5$\,\MLsun.

We take a slightly different approach in describing the spatial distribution of the three populations.
Instead of letting the Gaussian components in the surface brightness MGE each contribute an independent fraction to each of the three populations (as in Section \ref{sec:m1}), we adopt Legendre polynomials to describe the fractions of each stellar population as a function of radius.
While the former approach is maximally agnostic about the shape of the individual populations surface brightness distributions, it requires too many free parameters, not all of which contribute meaningfully to the fit.
In a two population model we require 10 free parameters, one for each MGE component, in a three population model, we would require 20 additional free parameters - 2 for each MGE component, which quickly becomes unfeasible.
By choosing Legendre polynomials we greatly reduce the number of additional parameters, at the expense of partly loosing the independence of the individual MGE fractions.
We believe, however, that this choice of parametrisation is still agnostic enough about shape of the populations' surface brightness distributions, so that we do not introduce involuntary bias to the fit.
We need two polynomials, as the fraction of the third population is determined by the remaining fraction from the total.
Similar to the functional description of the gravitational potential, we use the values of the Legendre polynomials at radii of $1\,\sigma$ of the MGE components to scale them accordingly.
We tested 2nd and 3rd order polynomials and found very similar results.
We chose to work with the 2nd order Legendre polynomials for simplicity.

In this model we also fit for the flattening, angular momentum, and anisotropy of the three populations independently, but we fixed the systemic velocity of M\,54 to the best fit value in the two population model to reduce the number of free parameters.

The three population model has $33$ free parameters: 4 parameters describing the gravitational potential ($\rm\Upsilon_0$, $\rm\Upsilon_t$, $\rm\Upsilon_{\infty}$, $\rm r_t$); 3 anisotropy ($\rm\beta_z^{YMR}$, $\rm\beta_z^{IMR}$, $\rm\beta_z^{OMP}$), 3 rotation ($\rm\kappa_{YMR}$, $\rm\kappa_{IMR}$, $\rm\kappa_{OMP}$), and 3 flattening ($\rm q_{YMR}$, $\rm q_{IMR}$, $\rm q_{OMP}$) parameters describing the kinematics of the three populations; 3 mean ages ($\rm \langle age\rangle_{YMR}$, $\rm \langle age\rangle_{IMR}$, $\rm \langle age\rangle_{OMP}$); 3 mean metallicities ($\rm\langle[Fe/H]\rangle_{YMR}$, $\rm\langle[Fe/H]\rangle_{IMR}$, $\rm\langle[Fe/H]\rangle_{OMP}$); 3 intrinsic metallicity spreads ($\rm\sigma_{[Fe/H]}^{YMR}$, $\rm\sigma_{[Fe/H]}^{IMR}$, $\rm\sigma_{[Fe/H]}^{OMP}$); 3 intrinsic photometric spreads (additional photometric uncertainty - $\rm\sigma_{V-I}^{YMR}$, $\rm\sigma_{V-I}^{IMR}$, $\rm\sigma_{V-I}^{OMP}$); 1 foreground parameter ($\epsilon$); 1 reddening parameter ($\rm E(B-V)$); 3 Legendre polynomial coefficients giving the shape and fraction of the YMR population surface brightness profile with respect to the adopted total surface brightness profile from \citet{noyola+gebhardt2006}; and 3 Legendre coefficients giving the shape and fraction of the IMR population surface brightness profile.
In Figure~\ref{fig:m2_corner} we show the posterior distributions of the main parameters of this population-dynamical model.
It is not possible to show all covariances in one figure, so we show the covariances between parameters pertaining to the Jeans equations and the isochrone fits separately, although it is a single unified model.
The best fit values of the main model parameters are summarised in Table \ref{tab:dyn_param}.
The posterior of the two Legendre polynomials coefficients, describing the radial fractions of the YMR and IMR populations, are shown in Figure~\ref{fig:mge_frac} in Appendix \ref{sec:app1}.

\begin{figure}
	\includegraphics[width=\columnwidth]{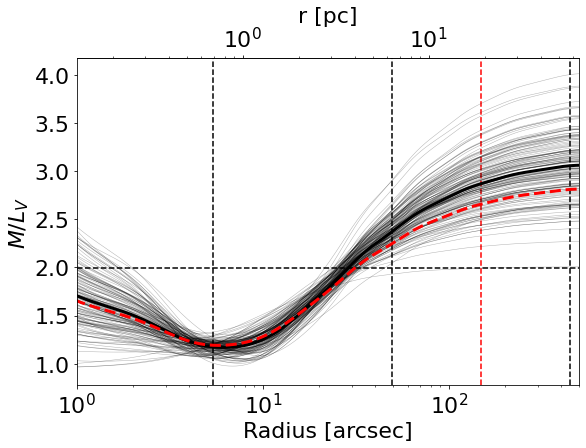}
    \caption{M\,54 best fit radial $\rm \Upsilon_V$ profile, according to the 3-population dynamical model described in Sect. \ref{sec:m2}, resulting from dividing the best fit mass and luminosity MGE models of the system. The median $\rm \Upsilon_V$ profile is indicated with a thick curve, while the thin curves represent random draws from the posterior distribution. The red dashed curve shows the median $\rm \Upsilon_V$ profile according to the 2-population dynamical model described in Sect. \ref{sec:m1.2}. The three black vertical dashed lines correspond to M\,54's core, half-light, and tidal radii \citep[GC catalogue of][2010 edition]{harris1996}, while the red vertical dashed line indicates the radial limit of our kinematic data. The horizontal dashed line corresponds to the average mass-to-light ratio for the entire system, e.g. the total mass divided by total luminosity out to the tidal radius.}
    \label{fig:m2_ml_var}
\end{figure}

We show the best fit $\Upsilon(r)$ profile for this model in Figure \ref{fig:m2_ml_var}.
It is very similar to what we found for the two population model qualitatively and quantitatively (Figure \ref{fig:m1_ml_var}).
With this model we estimate essentially the same total mass of Sgr's NSC as with the 2-population model with radially varying $\Upsilon$ - $1.60\pm0.07\times10^6$\,\Msun~out to its tidal radius ($\rm\langle\Upsilon_V\rangle=\frac{M_{tot}}{L_{tot}}=1.96\pm0.08$\,\MLsun), of which $65\%$ ($1.04\pm0.05\times10^6$\,\Msun) belongs to the OMP population, $20\%$ ($3.2\pm0.2\times10^5$\,\Msun) belongs to the YMR population, and $15\%$ ($2.4\pm0.2\times10^5$\,\Msun) belongs to the IMR population.
Table \ref{tab:m2_masses} gives the mass estimates and $\rm\Upsilon_V$ of the three populations locked within several key radial points.

\begin{table*}
	\centering
	\caption{Cumulative dynamical mass estimates and $\rm\Upsilon_V$ of M\,54's populations according to the model presented in Section \ref{sec:m2}}
	%\resizebox{\textwidth}{!}{
	\label{tab:m2_masses}
	\begin{tabular}{ccccccccc}
		\hline
        & OMP & IMR & YMR & Total & $\rm \Upsilon_{OMP}$ & $\rm \Upsilon_{IMR}$ & $\rm \Upsilon_{YMR}$ & $\rm \Upsilon_{Tot}$ \\
        &  [$10^6$\,\Msun]  &  [$10^6$\,\Msun] & [$10^6$\,\Msun]  &  [$10^6$\,\Msun]  & [\MLsun]  & [\MLsun]    &  [\MLsun]   & [\MLsun] \\	     
		\hline   
$\rm r_c$      & $0.07\pm0.01$ & $0.02\pm0.01$ & $0.04\pm0.01$ & $0.13\pm0.01$  & $1.38\pm0.06$ & $1.38\pm0.24$ & $1.37\pm0.10$ & $1.39\pm0.12$ \\
$\rm r_h$      & $0.56\pm0.02$ & $0.09\pm0.01$ & $0.22\pm0.01$ & $0.87\pm0.02$  & $1.65\pm0.04$ & $1.51\pm0.25$ & $1.51\pm0.10$ & $1.59\pm0.04$ \\
$\rm r_{fov}$ & $0.90\pm0.04$ & $0.15\pm0.01$ & $0.27\pm0.02$ & $1.32\pm0.04$  & $1.92\pm0.03$ & $1.78\pm0.24$ & $1.65\pm0.10$ & $1.83\pm0.06$ \\
$\rm r_t$       & $1.04\pm0.05$ & $0.24\pm0.02$ & $0.32\pm0.02$ & $1.60\pm0.07$  & $2.01\pm0.08$ & $2.21\pm0.17$ & $1.74\pm0.10$ & $1.96\pm0.08$ \\
		\hline
	\end{tabular}
	%}
\end{table*}

\begin{figure*}
	\includegraphics[width=\columnwidth]{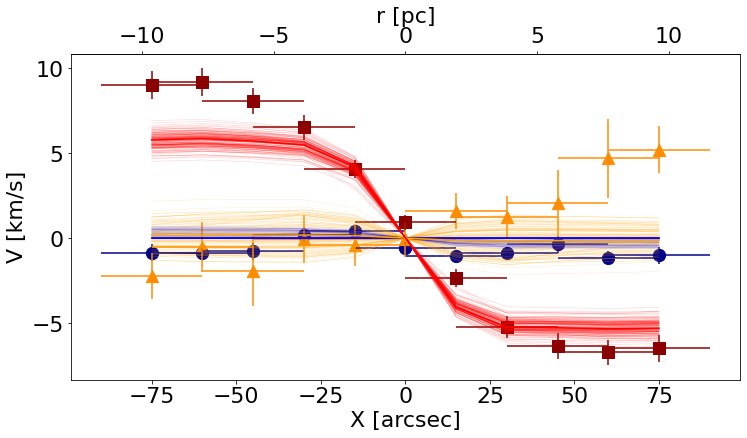}
	\includegraphics[width=\columnwidth]{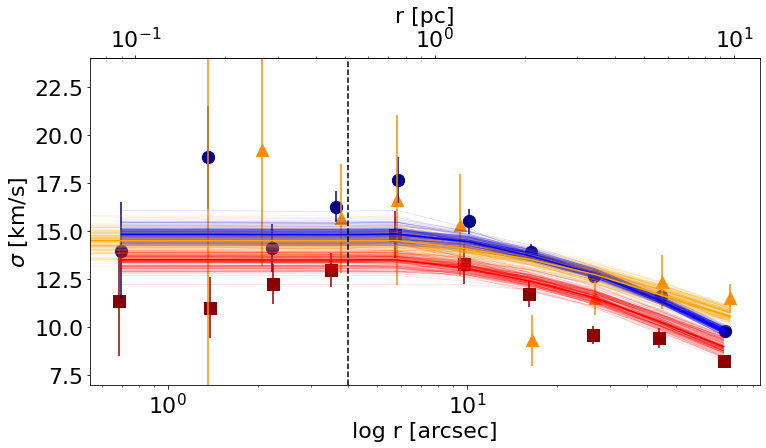}
    \caption{Predictions for the rotation ({\it left panel}) and the velocity dispersion ({\it right panel}) for the three population dynamical model (Sect. \ref{sec:m2}) drawn from the model posterior (YMR - red lines, IMR - green lines, OMP - blue lines), compared to the observed binned profiles (red, green, and blue dots with error bars, respectively), using stars with $>50\%$ probability to belong to either of the three populations. The horizontal error bars indicate the size of each bin. The vertical dashed line in the {\it right panel} shows the border between the NFM and WFM MUSE observations.}
    \label{fig:m2_kin}
\end{figure*}

\begin{figure*}
	\includegraphics[width=9cm]{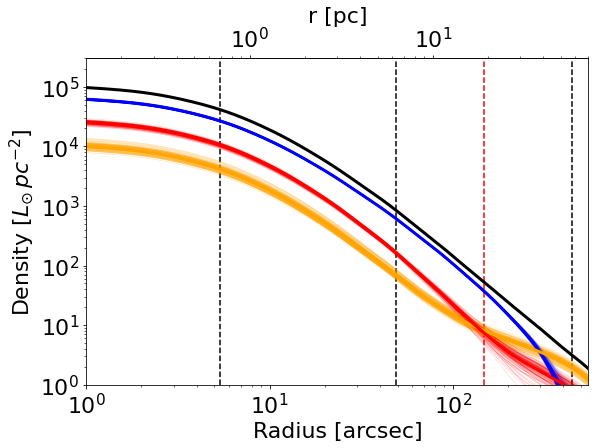}
	\includegraphics[width=9cm]{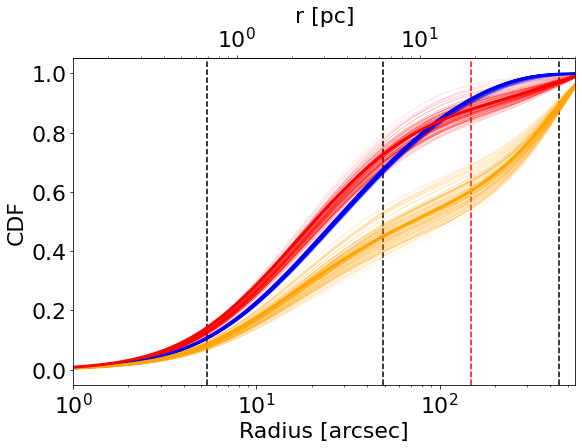}
	\includegraphics[width=9cm]{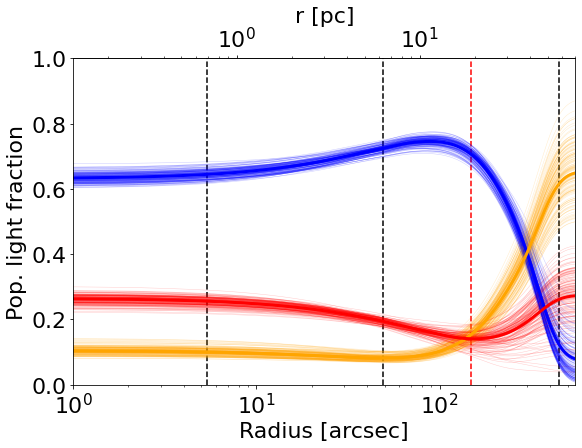}
	\includegraphics[width=9cm]{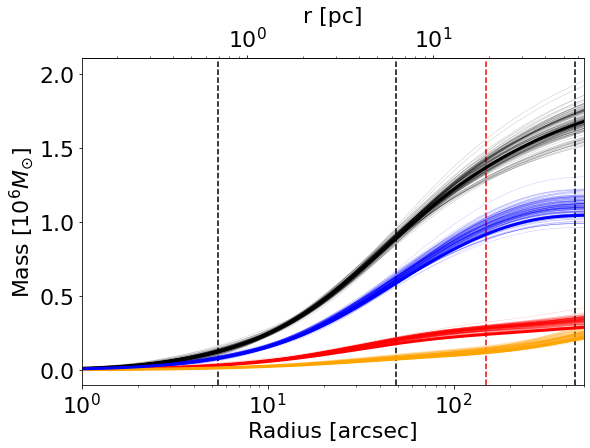}	
    \caption{{\it Top left panel:} Radial distribution functions of the three population-dynamical components of M\,54 (Sect. \ref{sec:m2}) as drawn from the model posterior (red lines for the YMR, green lines for the IMR, and blue lines for the OMP populations). The solid black line shows the combined surface brightness profile. {\it Top left panel:} Radial cumulative distribution function of the three population-dynamical components of M\,54. {\it Bottom left panel:}  Relative fractions of the three populations as function of radius. {\it Bottom right panel:} Cumulative masses of the three populations and the entire system (black lines). The black dashed lines indicate the core, half-light, and tidal radii of the system and the red dashed line indicates the limit of the MUSE field of view. 
    }
    \label{fig:m2_rdf_cdf}
\end{figure*}

The predictions of this model for the velocity dispersion and rotation profiles of the three different populations are shown in Figure~\ref{fig:m2_kin}.
According to the Jeans model, similar to the OMP, the IMR population has a negligible angular momentum, while the YMR population is the only one, which owes a significant fraction of its dynamical support to ordered motions.
The latter also has the lowest velocity dispersion.
It is interesting to note that there seems to be some evidence of counter rotation in the outer regions of the binned radial velocity profile of the IMR stars, which would put the Sgr dSph as another example of a galaxy with a kinematically decoupled core \citep{derijcke+2004,johnston+2018,fahrion+2019}.
The discrete Jeans model, however, has not picked up on this, predicting no rotation of the IMR population.
This could be due to limitations of the model as the fit is dominated by the denser central parts, where there appears to be no rotation.
We caution that due to the overall small number of stars associated with the IMR population, the uncertainties of the associated derived quantities are also considerably higher than for the YMR and OMP populations.
In addition, the stars with high probability of belonging to the IMR population have a relatively high velocity dispersion in the outer region of M\,54 and an overall flatter radial dispersion profile.
Our dynamical model does predict a flatter velocity dispersion profile for the IMR stars.%however there is still some small discrepancy between data and model predictions in the outer region of M\,54.

A flatter velocity dispersion profile of the IMR stars could be expected if they also have more extended density distribution and thus, the stars seen at small radii have preferably larger physical radii and appear close to the centre only in projection.
We found this indeed to be the case in Paper~I, and here we put this finding to a physical test with the population-dynamical model.
As mentioned above, we fit for the radial density profiles of the three population components within the dynamical model, where their radial luminosity fractions are described by 2nd order Legendre polynomials with the constraint that their sum is equal to the observed surface brightness profile of M\,54.
We show estimated radial density profiles of the three populations in Sgr's NSC and their respective cumulative functions, as well as the population fractions as a function of radius and their cumulative masses in Figure~\ref{fig:m2_rdf_cdf}.
We find that indeed the IMR population has the most extended surface density profile of the three, while the YMR populations remains the most centrally concentrated.
Our dynamical model predicts that the YMR population is slightly flattened ($q \simeq 0.85\pm0.04$) as expected from its higher degree of rotation, the OMP one ($q \simeq 0.93\pm0.04$) appears to be highly spherical with negligible degree of flattening, and the IMR falls in-between ($q \simeq 0.89\pm0.07$).

\begin{figure*}
	\includegraphics[width=8.3cm]{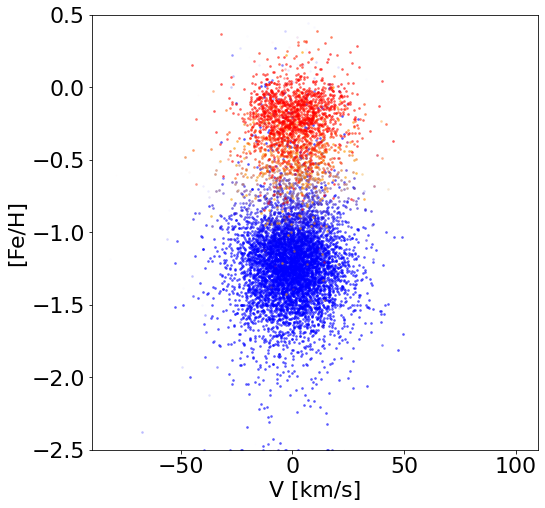}
	\includegraphics[width=8cm]{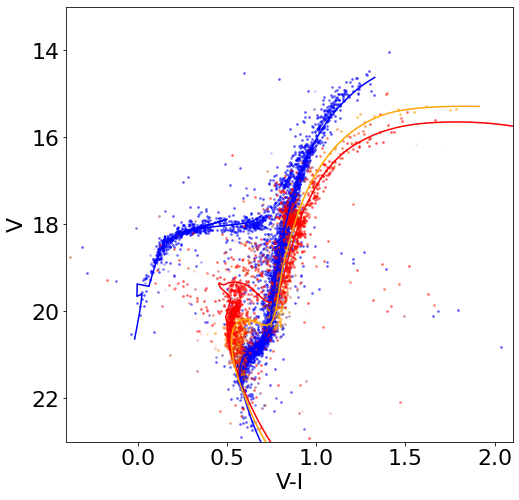}
	\includegraphics[width=6cm]{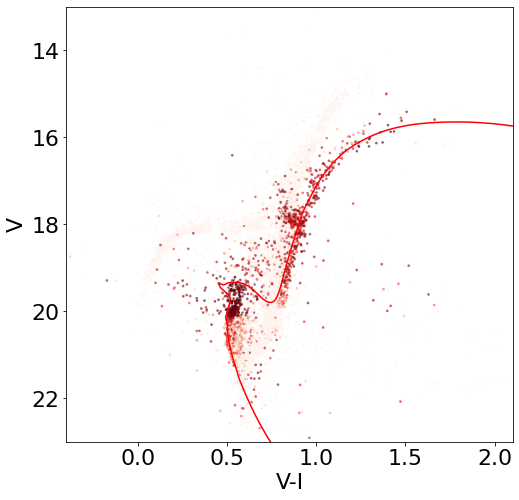}
	\includegraphics[width=6cm]{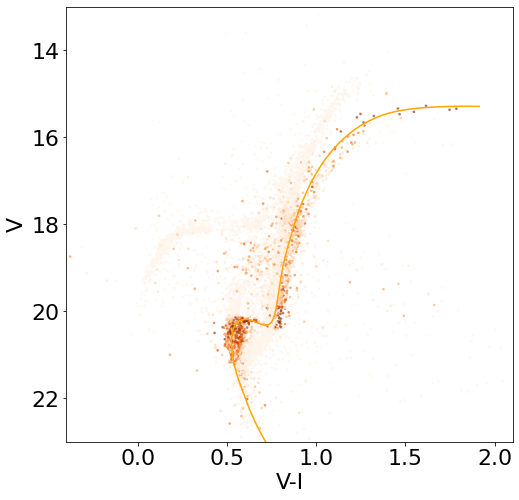}
	\includegraphics[width=6cm]{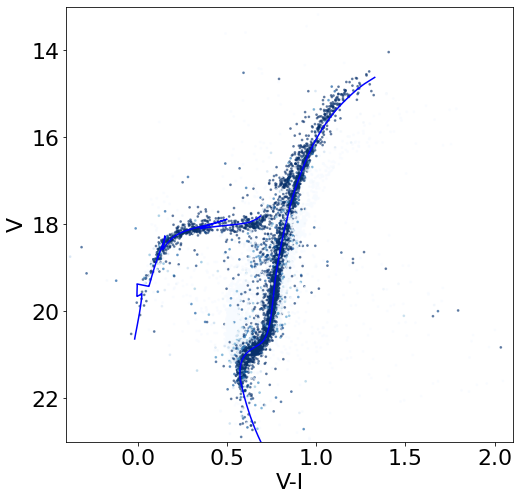}
    \caption{{\it Top:} Radial velocity vs. metallicity plot of all M\,54 member stars in the sample ({\it left panel}) and a CMD ({\it right panel}), colour-coded by the probability of the stars to belong to the YMR (red), the IMR (green) and the OMP (blue) populations according to the 3-population dynamical model (Sect. \ref{sec:m2}). {\it Bottom:} Separate CMDs of the observed stars, colour-coded according to the probability to belong to each of the three populations: YMR (shades of red), IMR (shades of green), and OMP (shades of blue). A darker colour indicates higher probability. The best fit Dartmouth isochrones are also plotted together with the HB model.}
    \label{fig:m2_pop}
\end{figure*}

In Figure~\ref{fig:m2_pop} we show how the stars of the three populations are separated probabilistically by our model on the CMD and plot the best fit isochrones.
We estimate the reddening towards M\,54 to be $\rm E(B-V) = 0.11$\,mag.
The YMR population has a mean age $2.53\pm0.01$\,Gyr, the IMR population $5.37\pm0.06$\,Gyr, and the OMP one $14.14\pm0.13$\,Gyr according to the Dartmouth isochrone models.
We find that the photometric uncertainties in \citet{siegel+2007} catalogue and the metallicity spreads appear sufficient to explain the CMD broadening, except for the YMR population, where we need an additional colour uncertainty of $\rm\sigma_{V-I}^{YMR} = 0.05$\,mag.
The median colour uncertainty of the photometric catalogue is $0.028$\,mag.

As for the metallicities, we find $\rm\langle[Fe/H]\rangle_{YMR} = -0.21\pm0.01$\,dex, with an intrinsic spread $\rm\sigma_{[Fe/H]}^{YMR} = 0.17\pm0.01$\,dex; $\rm\langle[Fe/H]\rangle_{IMR} = -0.47\pm0.01$\,dex, with an intrinsic spread $\rm\sigma_{[Fe/H]}^{IMR} = 0.26\pm0.02$\,dex; $\rm\langle[Fe/H]\rangle_{OMP} = -1.21\pm0.01$\,dex, with an intrinsic spread $\rm\sigma_{[Fe/H]}^{OMP} = 0.24\pm0.01$\,dex.
We find significant intrinsic metallicity spreads in all three populations and while part of these could be physical in nature for the IMR and OMP populations, it is difficult to explain a physical metallicity spread for the YMR population.
We note that $\rm\sigma_{[Fe/H]}^{YMR}$ is the lowest of the three populations and is likely dominated by systematic factors like underestimated metallicity measurement errors \citep[see Sect. 5.1 in][]{husser+2016}.

Note also that quoted uncertainties of the best fit parameters are purely statistical in nature and they are very low due to the large number of stars that are fitted simultaneously. 
The real uncertainties are dominated by systematic effects in the Dartmouth isochrone models, the photometry, the metallicity estimates, $\alpha$-enhancement, the assumed distance, etc., which exploration is beyond the scope of this work.

Overall, the results of the 3-population model in terms of M/L, rotation signatures and dispersion profiles are consistent with our previous model. 
However, the 3-component model can be used to better separate two MR populations in M\,54, which we find to have different spatial distribution, kinematics, mean ages, and mean metallicities.
They likely have different origin, where the IMR stars belong to the most central field population of Sgr, while the rapidly rotating YMR population is formed in an {\it in situ} burst of star formation in the nucleus of the dwarf galaxy.

\section{Discussion}\label{sec:discussion}

\subsection{Dynamical model comparison}

We presented three dynamical models in this work, exploring the effects of different gravitational potential parametrisation and changing the number of population components.
A summary of the estimated total mass and mass contributions of the different populations out to the tidal radius according to the three dynamical models is given in Table \ref{tab:bic}.

\begin{figure}
	\includegraphics[width=\columnwidth]{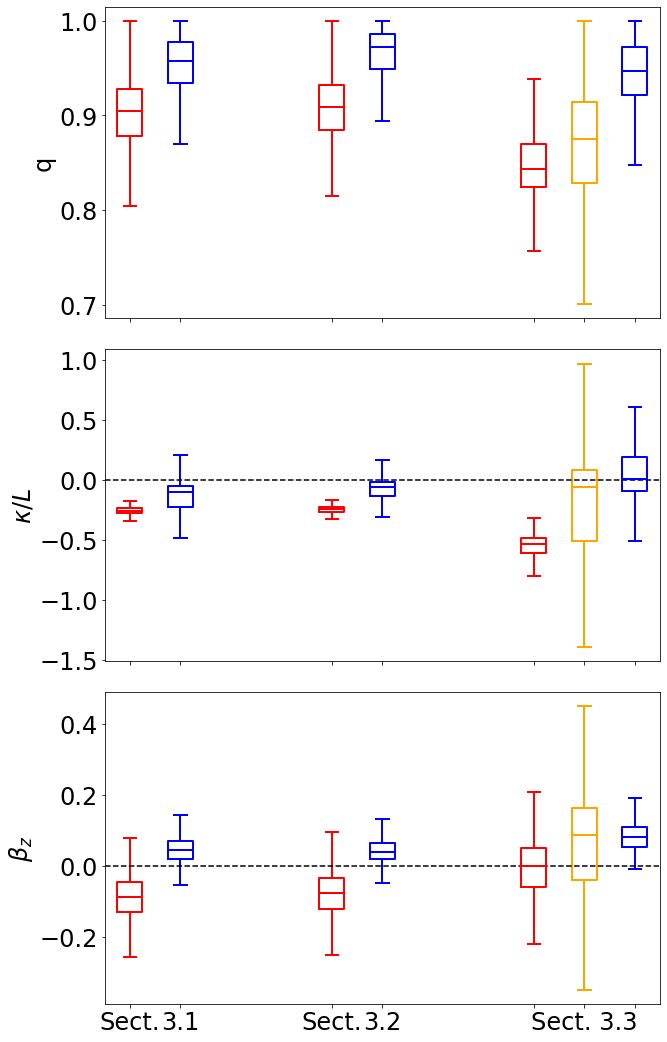}
    \caption{Comparison between the best fit values for the flattening  {\it (top panel)}, luminosity weighted rotation {in arbitrary units \it (middle panel)}, and anisotropy {\it (bottom panel)} of M\,54's modelled populations - YMR (red), IMR (orange), and OMP (blue), according to the three dynamical models described in Sect. \ref{sec:m1} (two populations with constant independent $\Upsilon_k$), Sect. \ref{sec:m1.2} (two populations with radially varying $\Upsilon(r)$), and Sect. \ref{sec:m2} (three populations with radially varying $\Upsilon(r)$), respectively.}
    \label{fig:par_comp}
\end{figure}

In Sect. \ref{sec:m1} and Sect. \ref{sec:m1.2} we considered two population components (MP \& MR), but changed the assumptions for the $\Upsilon$ relation - two separate constant mass-to-light ratios for the two populations (Sect. \ref{sec:m1}) vs. a common radially varying mass-to-light ratio (Sect. \ref{sec:m1.2}).
A problem of the former model is that it attributes almost all of the system's mass to a single population component.
Both models have significantly different predictions for the total mass of M\,54.

The dynamical model described in Sect. \ref{sec:m2} utilises the gravitational potential parametrisation with varying $\Upsilon$ and adds an additional population component.
It is fitted to a data set with expanded dimensions, because we included photometric information in addition to the metallicity.
The resulting total mass estimate for M\,54 is practically the same as in the two-population dynamical model with radially varying $\Upsilon$, which did not include photometric data in the fit (Table \ref{tab:bic}).
However, the three population dynamical model brings additional insight about the different origin of the metal-rich stars in Sgr's nuclear cluster, which couldn't be detected with a simpler model, mainly their different density distributions and angular momentum.

In Figure \ref{fig:par_comp} we compare the best fit values for the flattening, rotation, and anisotropy of the modelled populations in the three dynamical models.
There are no apparent discrepancies between the values of these parameters across the different models.
In all three models the metal rich stars are predicted to rotate faster, have a slightly flatter distribution, and be more tangentially anisotropic than the metal poor stars.
The latter, on the other hand, are more radially anisotropic.
One could also note that the separation of the YMR and IMR components from the metal rich population in the three population model, strengthens the dynamical differences between the most metal rich and metal poor stars in M\,54.
In the three population model, the YMR component has even higher angular momentum and ellipticity than in the two population model, where the YMR and IMR components are considered together.
Interestingly, in the three population model, the YMR population is consistent with pure isotropic rotation $\rm \kappa^{YMR} = -0.98\pm0.17$ and $\rm \beta_z^{YMR} = 0.00\pm0.08$ (see Table \ref{tab:dyn_param}).

In Table \ref{tab:bic} we compare the goodness of fit of the three models by looking at the median maximum likelihood ($\rm P_{max}$) reached by the MCMC fit, the number of free parameters ($\rm N_{free}$) of each model, and the Akaike and Bayesian information criteria (AIC \& BIC).
$\rm AIC = 2*N_{free} - 2*\ln P_{max}$ is a relative estimator of the quality of the fit, which penalises the model for the number of free parameters to avoid over-fitting.
Models with lower AIC values are preferred.
Similarly $\rm BIC = N_{free} * \ln(n) - 2*\ln P_{max}$ is an alternative estimator for the quality of the fit, which also takes into account the sample size (n).
In our case $\rm n = 8927$ is the number of stars with good quality velocity and metallicity measurements that we use to fit the dynamical models.
By its definition BIC penalises higher number of free parameters stronger than AIC.

It is evident from Table \ref{tab:bic} that the three population dynamical model formally provides a significantly better goodness of fit to the data, while the two population dynamical models with different gravitational potential treatment have comparable performance.
Note, however, that the two and three population model goodness of fit comparison is not entirely fair, because the fit was performed on different data sets (including and excluding photometric information).

\begin{table*}
	\centering
	\caption{Population-dynamical models information criteria, population mass fractions and total mass estimates at the tidal radius.}
	%\resizebox{\textwidth}{!}{
	\label{tab:bic}
	\begin{tabular}{ccccccccc}
		\hline
Model  & $\rm \ln P_{max}$ & $\rm N_{par}$ & AIC & BIC & $\rm M_{YMR}$ & $\rm M_{IMR}$ & $\rm M_{OMP}$ & $\rm M_{tot}$ \\
            &                               &                        &        &        &                           &                          &                         & $10^6$\,\Msun \\
		\hline   
Sect. \ref{sec:m1}    & $-43110$ & 24 & 86268 & 86438 & $4\%$   & ---                   & $96\%$ & $1.26\pm0.03$ \\
Sect. \ref{sec:m1.2} & $-43105$ & 26 & 86262 & 86447 & $33\%$ & ---                   & $67\%$ & $1.58\pm0.07$ \\
Sect. \ref{sec:m2}    & $-29380$ & 33 & 58826 & 59060 & $20\%$ & $15\%$          & $65\%$ & $1.60\pm0.07$ \\
		\hline
	\end{tabular}
	%}
\end{table*}

\subsection{Comparison with $N$-body models}

In this section we compare the results of our Jeans population-dynamical models with the findings by \citet{baumgardt2017}, based on $N$-body simulations, who find M\,54's dynamical mass to be $\rm M = 1.62\pm0.03\times10^6$\,\Msun ($\Upsilon_V = 2.18\pm0.20$).
This value is is in excellent agreement with our derived mass of the entire nuclear system in our three population model described in Sect. \ref{sec:m2} ($\rm M = 1.60\pm0.07\times10^6$\,\Msun; $\Upsilon_V = 1.96\pm0.08$) and for the two-population model with a radially varying mass-to-light ratio described in Sect.  \ref{sec:m1.2} ($\rm M = 1.58\pm0.07\times10^6$\,\Msun; $\Upsilon_V = 1.92\pm0.08$).
Some small differences between the two studies, like the lack of systemic rotation in the $N$-body model, the assumed distance to M\,54 ($23.5$\,kpc \citealt{baumgardt2017} vs. $26.5$\,kpc; this work) and its luminosity ($8.1\times10^5$\,\Lsun~\citealt{baumgardt2017} vs. $8.5\times10^5$\,\Lsun~this work) are enough to explain the small discrepancy in the mass-to-light ratio.

We only find a discrepancy between M\,54's mass estimates with respect to the $N$-body analysis for our dynamical model with constant mass-to-light ratios for the two populations (Sect. \ref{sec:m1}; $\rm M = 1.26\pm0.03\times10^6$\,\Msun; $\Upsilon_V = 1.51\pm0.04$), where we find a significantly lower figure.
This difference emphasises the importance for a correct parametrisation of the gravitational potential and shows that the radial variation of $\Upsilon$ cannot be ignored.

\subsection{A globular cluster and a nucleated dwarf}\label{sec:m3}

\begin{figure}
	\includegraphics[width=\columnwidth]{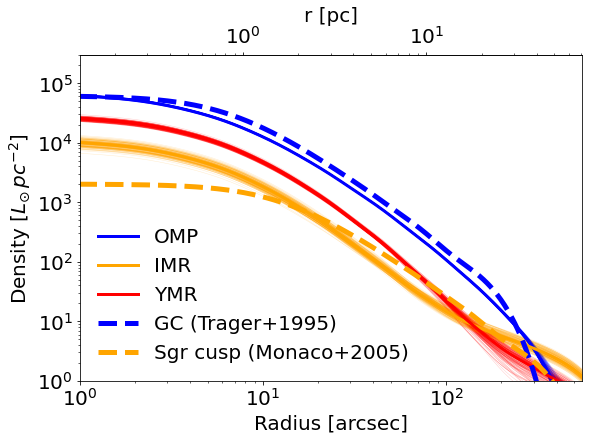}
    \caption{King profiles of the metal-poor stars \citep[blue line][]{trager+1995} and the metal-rich stars \citep[red line][]{monaco+2005}. The combined profile is plotted with a dashed black line and is compared to the surface brightness profile of \citet[][dashed green line]{noyola+gebhardt2006}.}
    \label{fig:monaco}
\end{figure}

In this section we discuss the results of our Jeans population-dynamical models in the context of historical population studies of Sgr's nuclear cluster.
\citet{monaco+2005} present the idea that the Sgr dwarf is a nucleated galaxy with a central over-density (cusp) of field stars that is coincident, but independent of the presence of a central GC.
They use a CMD to separate both stellar structures, which reflect the two dominant components of the Sgr NSC - the MP and MR stars.
Additional work by \citet{bellazzini+2008} and Paper~II show that the two populations also share a systemic velocity and thus are clearly colocated.
%Although the MP component has previously been referred to as a single GC, in Paper~I and in this work we showed that there is a clear metallicity spread, indicating that it is rather the combination of multiple old clusters that have merged in the centre of the galaxy.

\citet{monaco+2005} show that the surface brightness profile of the entire system can be well described with two \citet{king1962} profiles - the metal-poor GC stars follow a King profile with a core radius $r_c = 6.46\arcsec$ ($0.9$\,pc) and a tidal radius $r_t = 7.5\arcmin$ ($62$\,pc) from \citet{trager+1995}, while the metal-rich stars are significantly more extended with $r_c = 12.6\arcsec$ ($1.7$\,pc) and $r_t = 16.7\arcmin$ ($138$\,pc), and $\sim30$ times less dense in the centre than the GC population.

In Figure \ref{fig:monaco} we compare these two King profiles with our dynamical separation of three stellar structures - the YMR, IMR, and OMP populations.
Strictly speaking, our equivalent of the extended population from \citet{monaco+2005} would be the MR population from the two-population dynamical model (Sect. \ref{sec:m1.2}), which is also more spatially extended, but our predicted central density of the MR stars is slightly higher than the central density of the MP stars (Figure \ref{fig:m1_rdf_cdf}).
However, we have already shown in this work and in Papers~I~\&~II, that the metal-rich stars are not a homogeneous population, but rather at least two different stellar structures with different kinematics and origin.
In this case, the IMR stars from the three-population model (Sect. \ref{sec:m2}) are a better representation for the MR nucleus described by \citet{monaco+2005}.
Figure \ref{fig:monaco} shows that there is a generally good agreement between the density distributions of the MP stars from \citet{monaco+2005} and our OMP population.
The MR Sgr nucleus from \citet{monaco+2005} also follows a very similar density profile to our IMR population, except in the innermost $10$\,arcsec, where we find a significantly higher density.
The discrepancy could be due to lack of sufficient spatial resolution in the work by \citet{monaco+2005}.
We find the central density of the IMR stars to be five to six times lower than the central density of the OMP stars.

\subsection{The extended SFH of Sgr}\label{sec:m4}

The Sgr dSph has a very extended SFH \citep{deBoer+2015} and thus we expect that the field population also includes stars that have essentially the same ages and metallicities as the OMP NSC population.
However, in the population-dynamical models presented here we model the central over-density of Sgr field stars (the IMR component) as a single stellar population with a mean age and metallicity.
We do fit for its intrinsic metallicity spread and we find $\rm\sigma_{[Fe/H]}^{IMR} = 0.26\pm0.02$\,dex, the largest of the three populations.
The intrinsic metallicity spread of the OMP population is also significantly larger than expected for a GC ($\rm\sigma_{[Fe/H]}^{OMP} = 0.25\pm0.01$\,dex).
There are different possible explanations for the increased metallicity spread of the OMP stars - either this sample is contaminated by metal-poor Sgr field stars, or it is the post-merger remnant of multiple GCs with slightly different metallicities, or both.
In addition, part of the detected metallicity spread among the individual stellar populations could also be due to underestimated [Fe/H] measurement uncertainties, as discussed in \citet{husser+2016}.

In Paper~I we measured individual stellar ages and showed that the IMR stars have the largest intrinsic age spread ($1.16\pm0.07$\,Gyr), although this figure still does not include any metal-poor stars.
Here we chose to work with the more basic CMD information (magnitudes and colours), instead of relying on individual age estimates, which prevents us from deriving directly age spread estimates for the model-defined populations in M\,54.
However, we do not detect any additional broadening of the CMD than expected from the photometric uncertainties and metallicity spreads for either the OMP or IMR populations.
We do not expect to be able to distinguish between photometric broadening on the CMD caused by intrinsic age or metallicity spreads, as the effects are highly degenerate, especially at older ages.

Both the IMR and OMP populations appear to have very similar kinematic properties in our dynamical model (i.e. isotropic and non-rotating), hence we expect that we cannot discriminate between metal-poor field and GC stars kinematically.
If these populations also overlap in age and metallicity space, the only discriminating factor remains their different radial density distributions.
The Sgr field is indeed more radially extended than the GC population, considering the radial distribution of its metal-rich tail (the IMR population), however the OMP stars are significantly more abundant than the IMR stars at all radii in our data coverage, which makes it impossible to separate them in a meaningful way.

\section{Summary}\label{sec:summary}

This is the third paper of a series aiming to re-examine the massive star cluster M\,54 as Sagittarius dwarf galaxy's nuclear cluster.
In the first two papers of the series we published results based on the extensive MUSE seeing limited WFM mosaic data, obtained by our team during observing run 095.B-0585(A) (P.I.: L\"utzgendorf).
Our analysis then included $>7400$ individual stellar spectra with $\rm SNR>10\,px^{-1}$ extracted from the MUSE cubes with {\sc pampelmuse} \citep{kamann+2013}.
In this third instalment of the series we supplement the original stellar sample with additional WFM-AO and NFM-AO observations in the centre of the system, taken during the science verification of the AO capabilities of MUSE (P.I.: Alfaro-Cuello) for a total of $\sim9000$ unique stellar targets with $\rm SNR>10\,px^{-1}$ and improved spatial resolution in the inner region of M\,54.

In Paper~I we obtained spectroscopic metallicities and measured individual stellar ages from isochrone fits for the majority of the stars with extracted spectra.
We classified them in three distinct stellar populations that we called old metal-poor (OMP), intermediate-age metal-rich (IMR), and young metal-rich (YMR) and showed that they have different spatial distributions in the core of Sgr.
In Paper~II we explored the kinematic properties of these three stellar populations and showed that they exhibit different rotation and velocity dispersion profiles.
We offered a formation scenario for Sgr's NSC, which consists of stars with different origin and hence different population and kinematic characteristics: 
\begin{enumerate}[i)]%[label=(\roman*)]
\item a mixture of globular cluster stars (OMP) that belong to at least one massive, but possibly a merger of several globular clusters, due to their large metallicity spread;
\item stars formed recently {\em in situ} (YMR) that are very centrally concentrated and still have a high degree of rotation;
\item stars that belong to the inner Sgr field population (IMR) that have a large age spread, very extended spatial distribution, and a relatively flat velocity dispersion profile.
\end{enumerate}

In this work we explore the dynamical imprints of this formation scenario and the interplay between the different populations in a common gravitational potential.
We analysed all individual MUSE stellar spectra with {\sc spexxy}, utilising the {\sc phoenix} stellar library to derive radial velocities and metallicities.
This includes a re-analysis of the old sample with the slightly different methodology adopted in the current paper to get a truly uniform sample.
We use a dynamical modelling approach, based on the Jeans equations.
Each of the dynamical components in the model is characterised by its population properties (mean metallicity and age), spatial distribution, and velocity moments (rotation and velocity dispersion profiles), which we fit for simultaneously.
Our population-dynamical models are successful in estimating the joint probability of each star in the spectroscopic sample to belong to either one of the dynamical components of the model, or to the Milky Way foreground, based on its observed quantities (radial velocity, metallicity, photometry, and their respective uncertainties).

Overall we show that simple population-dynamical models, based on the Jeans equations, can explain the majority of observed properties of M\,54's complex stellar populations simultaneously and self-consistently.
We confirm our previous findings and especially emphasise that the metal-rich stars in M\,54 have heterogeneous origin.
The YMR population must have formed from in situ star-formation in a gaseous disk.
It is the most centrally concentrated, slightly more flattened than the rest, and with the highest angular momentum.
The IMR population corresponds to Sgr dSph's inner field population.
It is very extended spatially, follows a relatively flat velocity dispersion profile, and has a considerable metallicity spread.
Finally, the OMP population has all characteristics that are typical for globular clusters.
Such a mixed nuclear formation mechanism has been described in the literature \citep{neumayer+2011,denBrok+2014,antonini+2015,cole+2017,neumayer+2020,fahrion+2022}.

We also look into M\,54's mass and mass-to-light ratio ($\Upsilon$) radial profiles.
We conclude that taking into account the radial variations of the mass-to-light ratio is important to correctly reproduce the gravitational potential realistically.
We find that $\Upsilon_V$ is U-shaped with a minimum around the core radius, followed by a monotonic increase outwards (Figure \ref{fig:m2_ml_var}).
Multiple factors contribute to the complex $\Upsilon$ profile.
Mass segregated dark remnants and the centrally concentrated young stars have opposing effects to the mass-to-light ratio in the inner regions of the NSC, while the strong increase outwards could be due to an increased number of low mass stars and non-negligible dark matter contribution.
The total dynamical mass enclosed within the tidal radius ($76$\,pc) of the entire system and its different stellar components are summarised in Table \ref{tab:bic}.
Our results are in excellent agreement with $N$-body simulation studies by \citet{baumgardt2017}.

In a forthcoming work (Alfaro-Cuello et al., in prep.) we will address the question whether an intermediate-age black hole exists at the heart of M\,54, based on the dynamical models presented here and the work by \citet{aros+2020}.
These results provide a benchmark for studying the disruption of nucleated satellite galaxies and the formation of ultra-compact dwarfs.

{\bf Acknowledgements:}
We thank Holger Baumgardt for valuable discussions about the shape of M\,54's gravitational potential.
We thank Tim-Oliver Husser for providing the MUSE LSF corrected PHOENIX spectral library.
This research has made use of NASA’s Astrophysics Data System.
GvdV acknowledges funding from the European Research Council (ERC) under the European Union's Horizon 2020 research and innovation programme under grant agreement No 724857 (Consolidator Grant ArcheoDyn).
AMB acknowledges funding from the European Union’s Horizon 2020 research and innovation programme under the Marie Sk\l{}odowska-Curie grant agreement No 895174.

\appendix

\section{Corner plots for the individual populations surface brightness distribution parametrisation}\label{sec:app1}

\section{M\,54 complete dataset}\label{sec:app2}

\begin{table*}
	\centering
	\caption{M\,54 complete dataset used in this work. Here we show only the first three entries, the full table is available online. The stellar IDs, coordinates, and photometry are from the HST catalogue of \citet{siegel+2007}. The radial velocities and metallicities are from this work.}
	\resizebox{\textwidth}{!}{
	\label{tab:dataset}
	\begin{tabular}{cccccccccccc}
		\hline
ID        & RA                  & DEC                & $\rm V_r$ & Err.  $\rm V_r$ & [Fe/H]     & Err. [Fe/H] & F606W & F814W & Err. F606W & Err. F814W & Pointing      \\
            & [deg]              &  [deg]               & [\kms]       &  [\kms]              &                &                  & [mag]    & [mag]   &  [mag]          &  [mag]         &                    \\
		\hline   
36954 & 283.7653198 & $-$30.4947300 & 141.419   & 0.600                & $-$1.043 & 0.017        & 17.735 & 16.882 & 0.005            & 0.012          & WFM-noAO \\
41660 & 283.7614136 & $-$30.4950466 & 130.990   & 2.332                & $-$1.251 & 0.059        & 19.202 & 18.414 & 0.006            & 0.016          & WFM-noAO \\
40701 & 283.7621765 & $-$30.4926395 & 153.876   & 1.958                & $-$1.360 & 0.097        & 18.286 & 18.075 & 0.032            & 0.015          & WFM-noAO \\
		\hline
	\end{tabular}
	}
\end{table*}

\begin{figure*}
	\includegraphics[width=14cm]{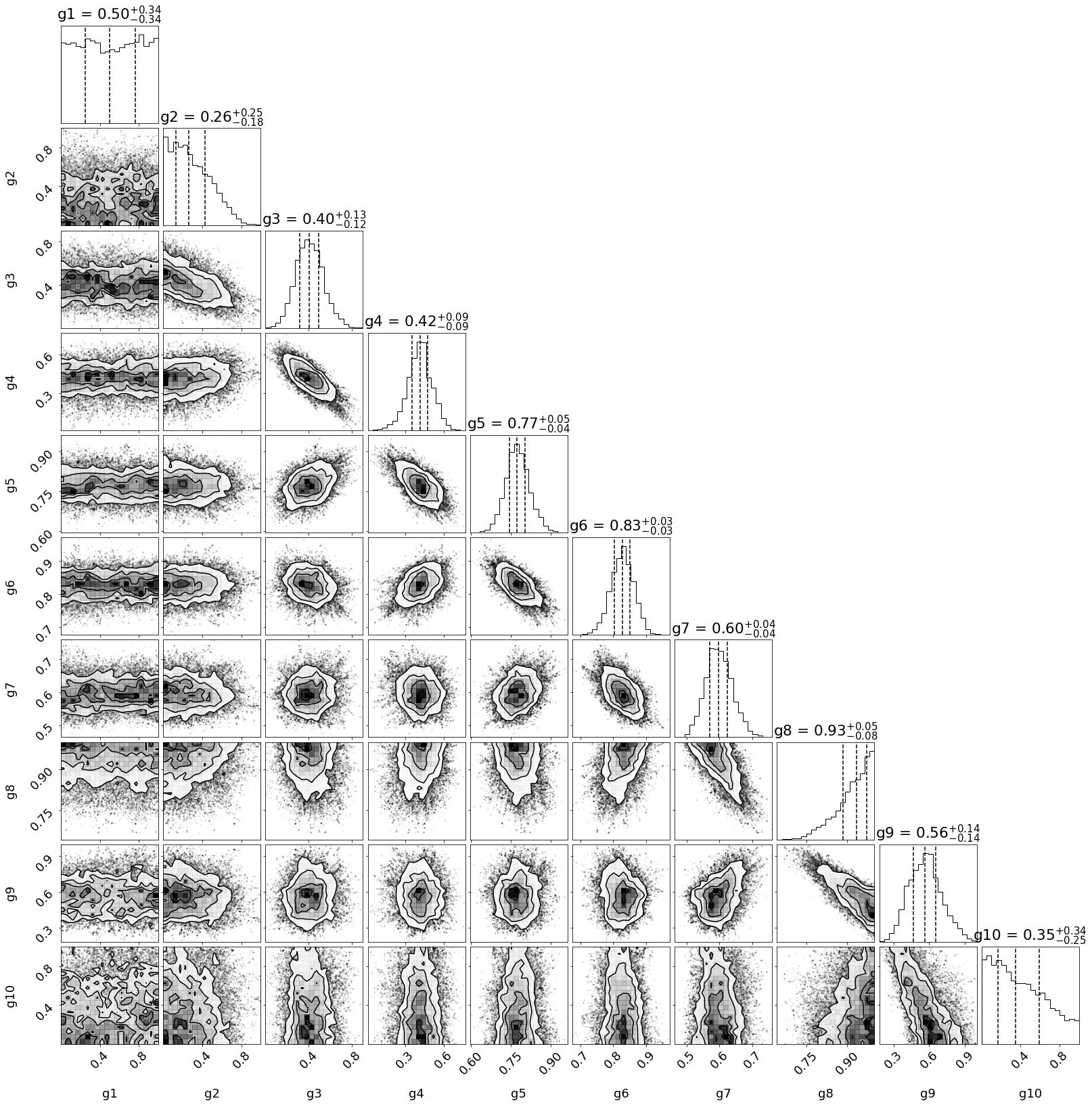}
	\includegraphics[width=7cm]{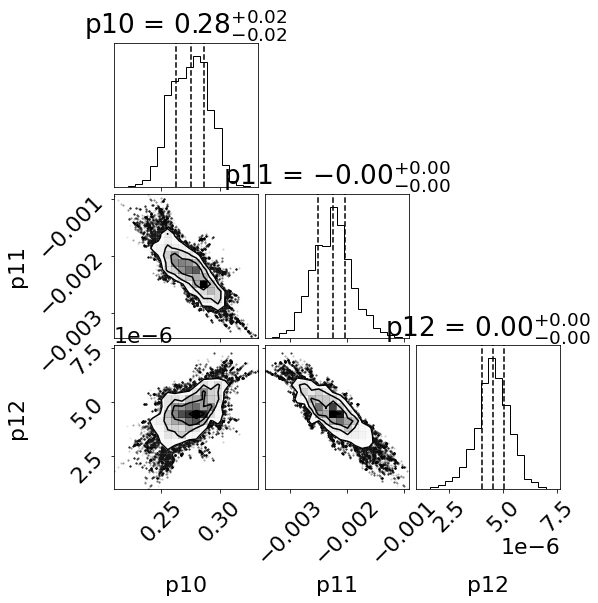}
	\includegraphics[width=7cm]{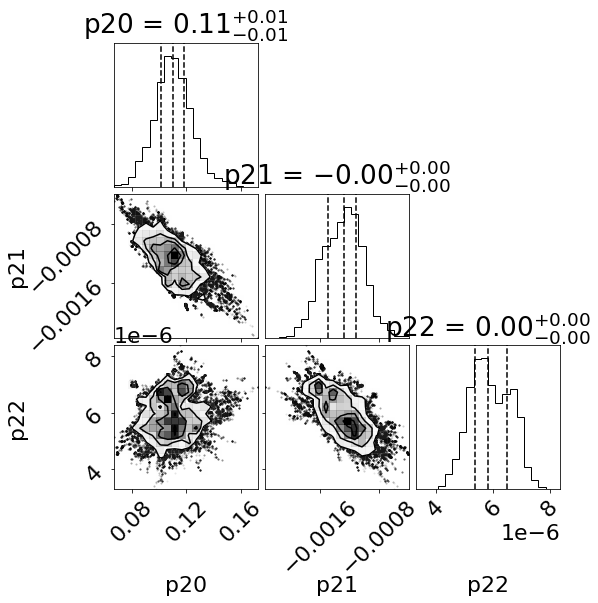}
    \caption{{\it Top:} corner plot of the individual fractions of the MGE components contributing to the MP population according to the best fit two population dynamical model presented in Sect. \ref{sec:m1}. {\it Bottom:} corner plot of the Legendre polynomial coefficients used to compute the YMR ({\it left}) and IMR ({\it right}) fractions of the surface brightness MGE in the three population dynamical model presented in Sect. \ref{sec:m2}.}
    \label{fig:mge_frac}
\end{figure*}


\bibliography{paper}{}
\bibliographystyle{aasjournal}



\end{document}